\documentclass{article}
\usepackage{graphicx}
\usepackage{amsmath,amssymb}
\usepackage{graphics,color,array,calc,rotating,epsfig,psfrag}
\usepackage{hyperref}
\numberwithin{equation}{section}
\usepackage{cite}
\usepackage{bm}
\usepackage{dcolumn}
\usepackage{float}
\usepackage{subfigure}
\usepackage{amsfonts}
\usepackage[mathscr]{eucal}
\def\be{\begin{equation}} \def\ee{\end{equation}}
\def\bea{\begin{eqnarray}} \def\eea{\end{eqnarray}}

\newcommand{\nn}{\nonumber}
\newcommand\Tstrut{\rule{0pt}{2.6ex}}       
\newcommand\Bstrut{\rule[-0.9ex]{0pt}{0pt}} 
\newcommand{\TBstrut}{\Tstrut\Bstrut} 
\begin{document}
\baselineskip 18pt%
\begin{titlepage}
\vspace*{1mm}%
\hfill%
\vspace*{15mm}%
\hfill
\vbox{
    \halign{#\hfil         \cr
          } 
      }  
\vspace*{20mm}

\begin{center}
{\large {\bf  Quasinormal modes for non-minimally coupled scalar fields in regular black hole spacetimes: Grey-body factors, Area spectrum and Shadow radius
}}\\
\vspace*{5mm}
{  Davood Mahdavian Yekta\footnote{d.mahdavian@hsu.ac.ir}, Majid Karimabadi, S. A. Alavi \footnote{s.alavi@hsu.ac.ir}}\\

\vspace*{0.2cm}
{$^{}$ Department of Physics, Hakim Sabzevari University, P.O. Box 397, Sabzevar, Iran}\\
\vspace*{1cm}
\end{center}
\begin{abstract}
 Black hole perturbation theory is a useful approach to study interactions between black holes and fundamental fields. A particular class of black hole solutions arising out of modification of Einstein's general theory of relativity are regular black holes (RBHs) which can be constructed using a nonlinear electrodynamic Lagrangian. Because of their importance, we are interested in studying the behavior of three kinds of such RBHs under perturbations generated by an external field. Indeed, we investigate the quasinormal modes (QNMs) of a massive scalar field propagating near the RBHs which is non-minimally coupled to the Ricci scalar tensor of background geometry. We will attempt to find the low-lying quasinormal frequencies of the perturbations by using WKB approximation. We shall also study the relationship between the QNMs and some characteristic properties of black holes such as grey-body factors, area quantization, and shadow radius.
\end{abstract}

\end{titlepage}
\section{Introduction}

Recent observations \cite{Abbott:2016blz,Abbott:2016nmj,Akiyama:2019cqa,Akiyama:2019eap} have provided us with the strong evidences that the black holes do exist. In this regard, the behavior of the fields surrounding a black hole not only tells us about its presence but also helps us to determine its parameters. After perturbation, a black hole undergoes damped oscillations with complex frequencies. The modes of such oscillations are called quasinormal modes (QNMs) which correspond to solutions of the wave equation satisfying the boundary conditions, in general appropriate for purely ingoing waves at the horizon and purely outgoing waves at the asymptotic infinity.
The study of QNMs of a black hole is an old and well established subject in physics \cite{Regge:1957td,Zerilli:1970se,Teukolsky:1972my}. Since there is a huge number of references on this subject, we only refer to some comprehensive reviews in Refs.~\cite{Kokkotas:1999bd,Berti:2009kk,Konoplya:2011qq}.

On the other hand, an old problem in the black hole physics is the presence of spacetime singularity at the center of black holes \cite{Schwarzschild:1916ae,Chandrasekhar:1931ftj,Penrose:1964wq}.
In this paper, we study the black holes without singularity known as regular black holes (RBHs) or singularity free black holes. The first kind of RBH space-time in general relativity was proposed by Bardeen \cite{Bardeen}, and shortly after revived in Refs.~ \cite{Borde:1994ai,Borde:1996df}. The Bardeen model satisfied the weak energy condition but it was not a vacuum solution of Einstein's equations, so it was necessary to introduce some external form of matter or a modification of gravity.  Ayon-Beato and Garcia (ABG) proposed a new nonlinear electrodynamics which, when coupled to gravity, produces an exact RBH solution that also satisfies the weak energy condition \cite{AyonBeato:1998ub,AyonBeato:1999ec,AyonBeato:2000zs}. Subsequently, further analyses of singularity avoidance have been proposed in Refs.~ \cite{Bronnikov:2000yz,Dymnikova:2004zc,Hayward:2005gi,Bronnikov:2006fu,Balart:2014cga,Nicolini:2005vd,Ansoldi:2006vg,Nicolini:2009gw}.

Recently, there has been a revival of interest in alternative theories including RBHs. Studying more realistic configurations, which include coupling terms are very important \cite{Ohashi:2004wr}. This provides strong motivations for investigating the coupling of a scalar field with the Ricci scalar tensor of RBH space-time geometry. There is a lot of interest in this kind of coupling in many research fields in physics, such as modified scalar-tensor theories (see e.g. \cite{Quiros:2019ktw} and Refs. therein), conformal gravity \cite{Mannheim:2011ds}, cosmological models \cite{Mannheim:2005bfa,Mannheim:2009qi,Faraoni:1996rf,Faraoni:2000wk}. In particular, we use WKB method to obtain the QNMs of perturbations of a scalar field with such a coupling around Bardeen, Hayward, and ABG black holes. The WKB method, initially proposed in Refs. \cite{Schutz:1985zz,Iyer:1986np,Iyer:1986nq} to obtain the QNMs of Schwarzschild background, provides a simple and powerful tool for studying properties of black holes. Further calculations for the Kerr and Reissner-Nordstrom black holes can be found in Refs. \cite{Kokkotas:1988fm,Kokkotas:1991vz,Konoplya:2003ii,Zhidenko:2003wq}. There are other analytical and numerical methods than WKB to compute the QNMs of a particular black hole perturbations in Refs.~\cite{Vishveshwara:1970zz}-\!\cite{Panotopoulos:2020mii}.

Using the WKB formula, the QNMs of neutral and charged minimal scalar field perturbations for RBHs have been studied in Ref.~\cite{Flachi:2012nv} and for wormholes and RBH using a phantom scalar field in Ref.~\cite{Bronnikov:2012ch}. Other calculations about the QNMs of RBHs in the case of minimal couplings could be found in Refs.~\cite{Fernando:2012yw,Toshmatov:2015wga,Breton:2016mqh,Fernando:2016ksb,Panotopoulos:2019qjk}  Perturbations of gravitational and Dirac fields in these backgrounds have also been studied in Refs.~\cite{Ulhoa:2013fca,Saleh:2018hba,Li:2013fka,Wahlang:2017zvk,Dey:2018cws,Chakrabarty:2018skk}.
However, these attempts are for theories with no coupling to the geometry, thus the main aim of this paper is to study the QNM spectra due to this kind of coupling. The QNMs are also important in determining some characteristic properties of black holes such as grey-body factors (GF), area quantization (AQ), and shadow radius (SR). These parameters are very important in analysis and verification of the obtained results from \cite{Abbott:2016blz,Abbott:2016nmj,Akiyama:2019cqa,Akiyama:2019eap}. This fact strongly motivates us to investigate the relationship between the QNMs and these quantities in the case of RBHs.

Very long ago, Hawking showed that a black hole can emit thermal radiation if the quantum effects are considered, known as Hawking radiation (HR) \cite{Hawking:1974sw,Hawking:1974rv,Hawking:1982dh}. Because of the non-trivial spacetime geometry near the black hole, the initial radiation received by a distant observer will get modified by a coefficient called grey-body factor \cite{Das:1996we,Brady:1996za,Gubser:1997yh}. We will apply a numerical prescription employing WKB approximation based on \cite{Konoplya:2009hv,Konoplya:2010kv,Toshmatov:2016bsb} to study the behavior of GFs for RBHs in different regimes of parameters. The are also some efforts on the calculation of GFs in coupling of scalar theory with matter fields \cite{Panotopoulos:2017yoe}, four-dimensional Gauss-Bonnet gravity in \cite{Konoplya:2020cbv}, and 2+1-dimensional BTZ black holes in \cite{Panotopoulos:2016wuu,Rincon:2018ktz,MahdavianYekta:2018lnp,Ovgun:2018gwt}.

It was conjectured by Bekenstein \cite{Bekenstein:1973ur,Bekenstein:1974jk} that the black hole entropy should be represented by a discrete spectrum in Planck units.
That is, the area of a classical black hole behaves like an adiabatic invariant, and so, according to Ehrenfest’s theorem, the corresponding quantum operator must have a discrete spectrum. Moreover, Hod proposed \cite{Hod:1998vk} an equally spaced area spectrum and used the apparent existence of a unique QNM frequency in the large damping limit to uniquely fix the spacing. However, Maggiore \cite{Maggiore:2007nq} suggested that a black hole can be viewed as a damped harmonic oscillator whose physically relevant frequency is identical to the complex QNM frequencies having both real and imaginary parts. According to \cite{Maggiore:2007nq}, for the highly excited QNMs the imaginary part is dominant over the real part and one can compute the AQ spectrum from imaginary part. We employ this prescription and a near horizon approximation to find equally spaced area spectrum in the case of RBHs in this paper.

Another important phenomenological feature of a black hole which is also closely related to the QNMs, is its SR \cite{Cardoso:2008bp}. The shadow image of the supermassive black hole in the center of $M87^{\ast}$ galaxy released by EHT \cite{Akiyama:2019cqa,Akiyama:2019eap} greatly stimulated our enthusiasm for the research of RBH's SR. More importantly, the research in this direction will open a new window to study the strong gravitational region near black hole horizon.
The investigation of SR for several black holes have been done in Refs.~\cite{Hioki:2008zw,Hennigar:2018hza,Grenzebach:2014fha,Wei:2019pjf,Konoplya:2019fpy,Jusufi:2020cpn,Liu:2020ola}. There are also some attempts to compute SR of different non-singular black holes in Refs~.\cite{Abdujabbarov:2016hnw,Amir:2016cen,Jusufi:2020agr}

The organization of the paper is as follows: In Sec.~2, we study the RBHs as solutions to the field equations of a non-linear electromagnetic model, which in the weak field limit are reduced to the standard Maxwell's linear theory. In Sec.~3, we introduce the non-minimal coupling model and discuss the effective potential appeared in the Schr\"{o}dinger-like equation of the scalar field dynamics. In Sec.~4, we use WKB method to derive the QNMs spectra and investigate the effects of physical parameters on the imaginary and real parts of frequencies. We provide a numerical verification on the GFs of HR for different values of RBH's parameters. We also determine the AQ spectrum from the near horizon consideration and SR in the eikonal limit for this family of black holes. Finally, Sec.~5 is devoted to a brief summary and concluding remarks.
\section{Regular black holes in non-linear electrodynamics}
The Bardeen model \cite{Bardeen}, as the first RBH model in general relativity, is reinterpreted as the gravitational field of a non-linear magnetic monopole, i.e., as a magnetic solution of Einstein's field equations coupled to a non-linear electrodynamics \cite{AyonBeato:1998ub,AyonBeato:2000zs,AyonBeato:1999ec}. The model is described by the action
\be \label{act} S=\int d^4 x\,\sqrt{-g}\left(\frac{1}{16\pi G}\,R-\frac{1}{4\pi} \,\mathcal{L} (F)\right),\ee
where $R$ is the scalar curvature and the Lagrangian of non-linear electrodynamics $\mathcal{L} (F)$ as a function of $F=\frac14 F_{\mu\nu} F^{\mu\nu}$, is given by
\be\label{NLE}\mathcal{L} (F)=\frac{3}{2\alpha q^2}\left(\frac{\sqrt{2q^2 F}}{1+\sqrt{2q^2 F}}\right)^{\frac52}.\ee
Assume that $G=1$ and $F_{\mu\nu}=\nabla_{\mu}A_{\nu}-\nabla_{\nu}A_{\mu}$. The parameter $\alpha\equiv \frac{|q|}{2M}$ is defined from the charge and the mass of the black hole. The regularizing parameter $q$ can be physically interpreted as the monopole charge of a self-gravitating magnetic field of non-linear electrodynamics \cite{AyonBeato:1998ub}.

The field equations of motion derived from (\ref{act}) are given by
\be\label{eom}
{G_{\mu}}^{\nu}=2(\mathcal{L}_{F} \,F_{\mu\rho}F^{\nu\rho}-{\delta_{\mu}}^{\nu} \mathcal{L}_{F}),\qquad \nabla_{\mu}(\mathcal{L}_{F} \,F^{\rho\mu})=0,\ee
where $G_{\mu\nu}$ is the Einstein's tensor and $\mathcal{L}_{F}\!=\!\mathcal{L} (F)$. In general, a static spherically symmetric RBH solution is given by
\be\label{bm} ds^2=-f(r) dt^2+\frac{dr^2}{f(r)}+r^2 (d\theta^2+ \sin^2\theta d\varphi^2)\,,\ee
where the function $f(r)$ is denoted by
\be \label{func} f(r)=1-\frac{2 m(r)}{r}\,.\ee
In this paper we are going to study three families of RBH solutions;
\begin{itemize}
\item Bardeen\cite{Bardeen},
\be \label{Bardeen} m(r)=\frac{Mr^3}{(r^2+q^2)^{3/2}}\,,\ee
\item Hayward\cite{Hayward:2005gi},
\be \label{Hayward} m(r)=\frac{Mr^3}{(r^3+q^3)}\,,\ee
\item ABG\cite{AyonBeato:1998ub},
\be \label{ABG} m(r)=\frac{Mr^3}{(r^2+q^2)^{3/2}}-\frac{q^2 r^3}{2(r^2+q^2)^2}.\ee
\end{itemize}

In the asymptotic limit $r\rightarrow \infty$, (\ref{Bardeen}) and (\ref{Hayward}) behave as Schwarzschild black hole of mass $M$, while (\ref{ABG}) looks like a Reissner-Nordstr\"{o}m (RN) black hole of mass $M$ and charge $q$. In the limit $r\rightarrow 0$, All of them behave as de Sitter spacetime \cite{Toshmatov:2015wga}.
The magnetic monopole charge is obtained from field strength tensor $F_{\theta\varphi}=q\sin\theta$ as
\be\frac{1}{4\pi} \int_{S_2^{\infty}} F=\frac{q}{4\pi}\int_{0}^{\pi}\int_{0}^{2\pi}\sin\theta\, d\theta d\varphi=q. \ee

\begin{figure}[H]
\centering
\subfigure[Bardeen]
{\includegraphics[width=.48\textwidth]{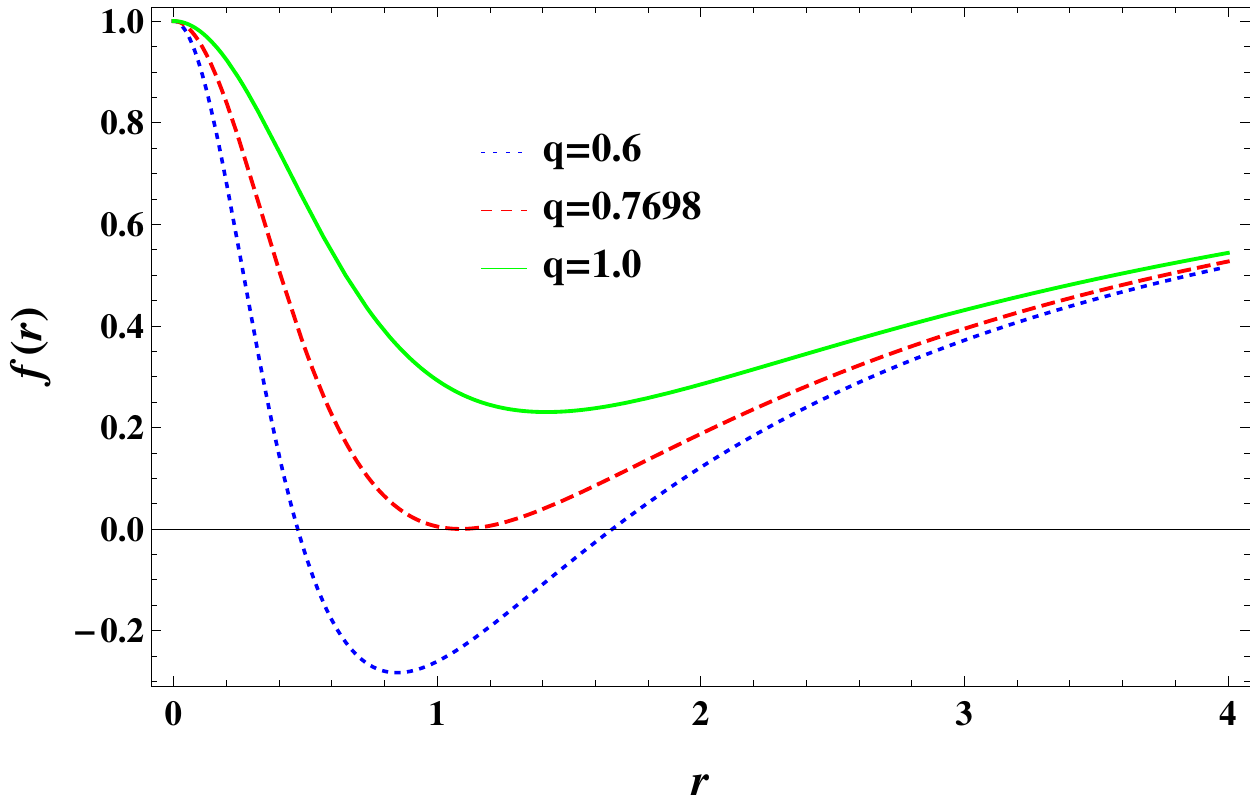}}
\subfigure[Hayward]
{\includegraphics[width=.48\textwidth]{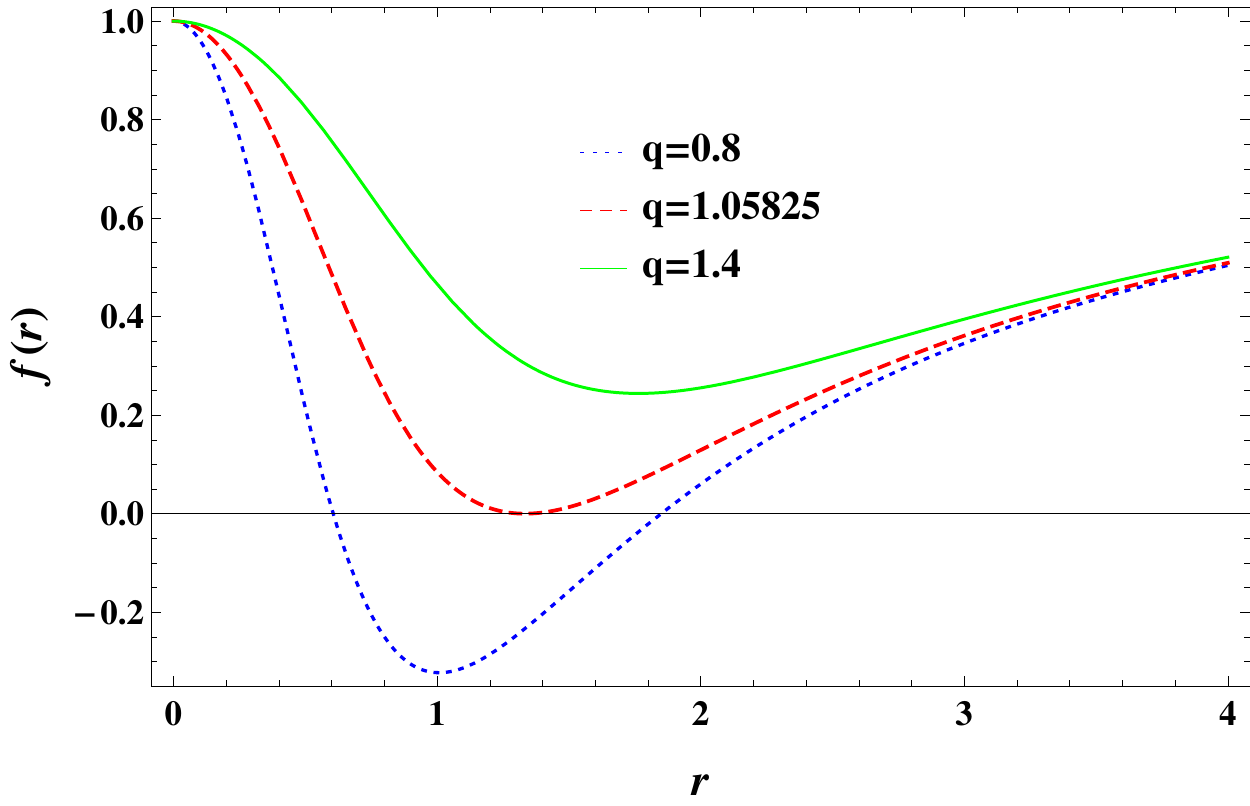}}
\subfigure[ABG]
{\includegraphics[width=.48\textwidth]{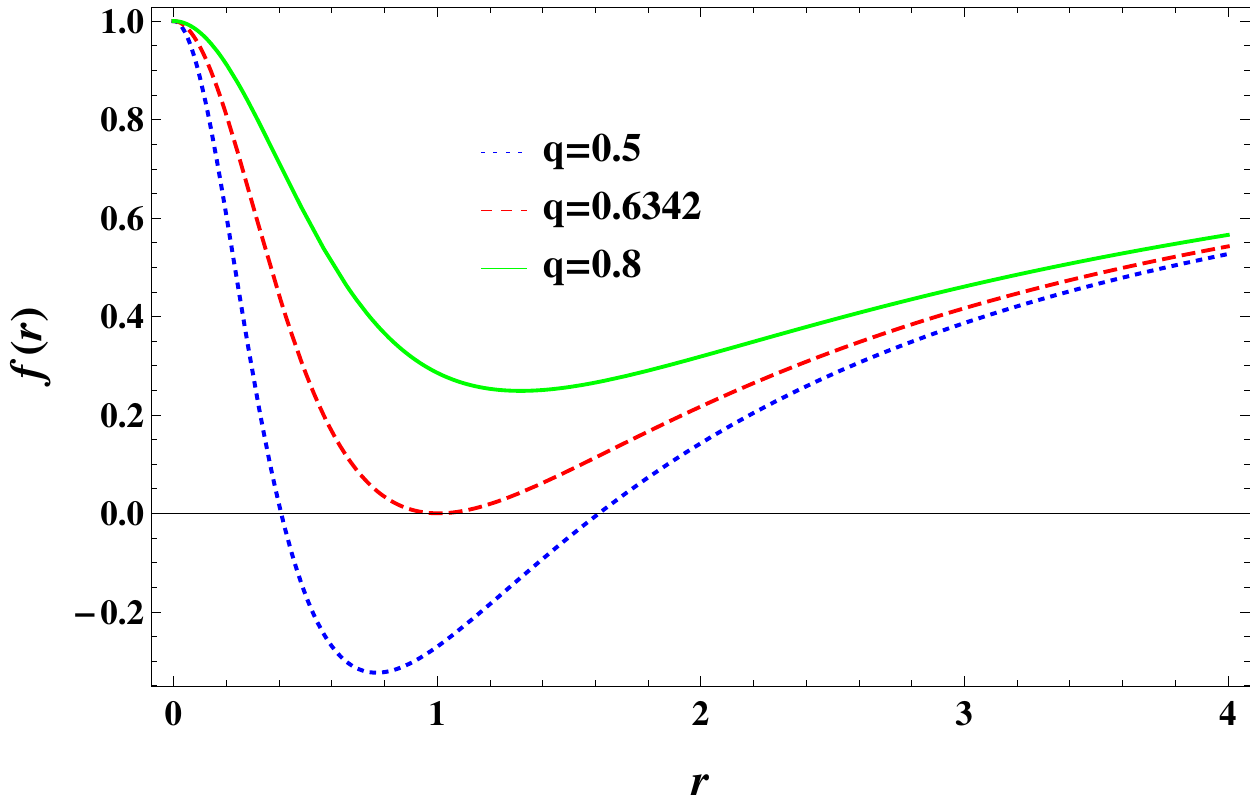}}
	    \caption{{ Radial behavior of the lapse function $f(r)$ for different values of $q$ when $M\!=\!1$.}}
\label{f1}
\end{figure}

The regularity of these solutions can be realized by calculating the curvature invariants, which for instance we carried out them for Hayward black hole in the following expressions
\bea\label{invs}
R_{\mu\nu\rho\lambda}R^{\mu\nu\rho\lambda}&\!\!\!=\!\!\!&\frac{48M^2(2q^{12}-2r^3q^9+18r^6q^6-4r^9q^3+r^{12})}{(r^3+q^3)^6}\quad \xrightarrow{r\rightarrow 0}\quad \frac{96M^2}{q^6}\,,\nn\\
R_{\mu\nu}R^{\mu\nu}&\!\!\!=\!\!\!&\frac{72M^2q^6(2q^6-2r^3q^3+5r^6)}{(r^3+q^3)^6}\quad \xrightarrow{r\rightarrow 0}\quad \frac{144M^2}{q^6}\,,\\
R&\!\!\!=\!\!\!&\frac{12Mq^3(2q^3-r^3)}{(r^3+q^3)^3}\quad \xrightarrow{r\rightarrow 0}\quad \frac{24M}{q^3}.\nn
\eea
All of them are regular everywhere and one can check that, this is also true for other solutions. The location of the horizons of the static spherically symmetric black holes are determined by solving the equation $f(r)=0$. Depending on the value of $q$, they may have no horizon at all, one degenerate horizon, or two distinct horizons. In order to clarify this concept we have depicted the metric function $f(r)$ in Figs.~(\ref{f1}) for the corresponding three RBHs. We call the case of degenerate horizon as the extremal solution, however this case is very different from general definition of RN black holes where $|q|\!=\!M$. The extremal values of the charge for Bardeen, Hayward, and ABG RBHs are respectively given by $q_{B}\!\sim\!0.7698$,  $q_{H}\!\sim\!1.0582$, and  $q_{A}\!\sim\!0.6342$.

\section{Propagation of a massive scalar field near RBHs}

In order to gain insight into the quantum nature of RBHs, the kinematical properties provide relevant clues about their semiclassical aspects. From theoretical point of view, there are two different ways to initiate the perturbation of a black hole; one is by adding external test fields to the black hole geometry and the other is by perturbing the black hole metric itself or gravitational perturbation. The simplest way to study black hole perturbations due to external fields is to study the scalar wave equation in the vicinity of a black hole geometry. Therefore, we consider the perturbations of a massive scalar field near the RBHs which is non-minimally coupled to the Ricci scalar tensor of the background geometry. But due to the linear approximation, we ignore the back-reaction of the scalar field on the background geometry and the field equation has a general covariant form in the RBH background.

The action governing the dynamics of this scalar field is given by
\be\label{sact} S=-\frac12 \int d^4 x\,\sqrt{-g}\,\left((\nabla \phi)^2+\mu^2 \phi^2+\xi R \phi^2\right),\ee
where $\xi$ is a dimensionless coupling constant and $\mu$ is the scalar mass. Also $g_{\mu\nu}$ and $R$ are respectively the metric and Ricci scalar of the RBHs geometries described by (\ref{bm})-(\ref{ABG}). Varying the action (\ref{sact}) with respect to $\phi$ gives a modified Klein-Gordon equation for the scalar field as
\be\label{eomsc} [\Box -\mu^2-\xi R]\,\phi(t,r,\theta,\varphi)=0.\ee

In the case of $\mu\!=\!0$ and $\xi\!=\!\frac16$ we have conformal coupling for which the scalar
field theory becomes conformally invariant \cite{Mannheim:2011ds}. This theory is able to tackle the problems of Dark Matter, Dark Energy and quantum gravity \cite{Mannheim:2005bfa,Mannheim:2009qi}. It should be noted again that we treat the scalar field as a weak external field that probes the black hole. We introduce the scalar field by the following standard ansatz with factorization
\be \label{ansatz} \phi(t,r,\theta,\varphi)=R(r) Y_{lm}(\theta,\varphi)\, e^{-i \omega t}\,,\ee
where $Y_{lm}(\theta,\varphi)$ denotes the spherical harmonic function
of degree $l$ satisfying $\Box_{(\theta,\varphi)} Y_{lm}\!=\!-l(l+1)Y_{lm}$ and $\omega$'s are the QNM frequencies of scalar perturbations. As mentioned earlier, these frequencies are complex and can be generally expressed in the form $\omega\!=\!\omega_{R}+i\omega_{I}$, so the factor $e^{-i\omega t}$ as the time-dependence of scalar field becomes $e^{-i\omega_{R} t}\,e^{\omega_{I}t}$. Therefore, the real part determines the normal frequency of the oscillations, i.e. $\frac{\omega_{R}}{2\pi}$, while the imaginary part represents the damping time of the mode as $t_{D}^{-1}\!=\!|\omega_{I}|$. In fact, the mode is unstable (exponentially growth) when $\omega_{I}>0$ and stable (exponentially decay) when $\omega_{I}<0$. In other words, the black hole is stable under dynamical perturbation since the scalar field vanishes as time passes for $\omega_{I}<0$.

If we redefine the radial function as $R(r)\!=\!\frac{\psi(r)}{r}$, then the radial part of the master equation (\ref{eomsc}) changes to the following equation
\be \label{req} -f(r) \frac{d}{dr}\left[f(r) \frac{d}{dr} \psi(r)\right]+V(r) \psi(r)=\omega^2 \psi(r)\,,\ee
where $V(r)$ is a nontrivial function of $r$ and other parameters in the model. By introducing a standard tortoise coordinate
\be\label{tc} dr^{*}=\frac{dr}{f(r)},\ee
we obtain a Shr\"{o}dinger-like wave equation (Regge-Wheeler wave-like equation \cite{Regge:1957td}) for the perturbation of RBHs by a scalar field as follows
\be \label{sceq}\frac{d^2 \psi}{{dr^{*}}^2}+(\omega^2-V(r)) \psi=0.\ee

Now, the function $V(r)$ is an effective potential given by
\be \label{veff} V(r)=-\frac{f(r)}{r^2}\left[r^2 \xi f''(r)+(4\xi-1) r f'(r)-\mu^2 r^2+2\xi f(r)-2\xi-l(l+1)\right],\ee
where the ``prime" standing for the derivative with respect to $ r$ and $f(r)$ is given in (\ref{func}) for different regular backgrounds (\ref{Bardeen})-(\ref{ABG}). In general, the radial solutions of equation (\ref{sceq}) are determined by some particular boundary conditions at $r^{*}\!=\!\pm\infty$ which are correspond to the physical observers at infinity and near the event horizon, respectively\cite{Kokkotas:1999bd}. For asymptotically flat spacetimes which we consider here, these conditions lead to
\be \psi \sim e^{-i\omega(t+r^{*})}\,\ee
at the event horizon $r^{*}\rightarrow -\infty (r\rightarrow r_{h})$, and
\be \psi \sim e^{-i\omega(t-r^{*})}\,\ee
near the asymptotic infinity $r^{*}\rightarrow +\infty (r\rightarrow \infty)$. The first solution that means the wave is purely ingoing at the horizon expresses the fact that nothing escapes from the horizon, on the other hand the second purely outgoing wave corresponds to the requirement that no radiation comes from infinity. Of course in another prescription, the observers can also detect an incoming radiation at infinity which is important in determining the resonant QNMs of the black hole, a mode whose response to an external perturbation is maximum \cite{Iyer:1986np,Dappiaggi:2018xvw}.

\begin{figure}[H]
\centering
\subfigure[\scriptsize{Bardeen}]
{\includegraphics[width=.49\textwidth]{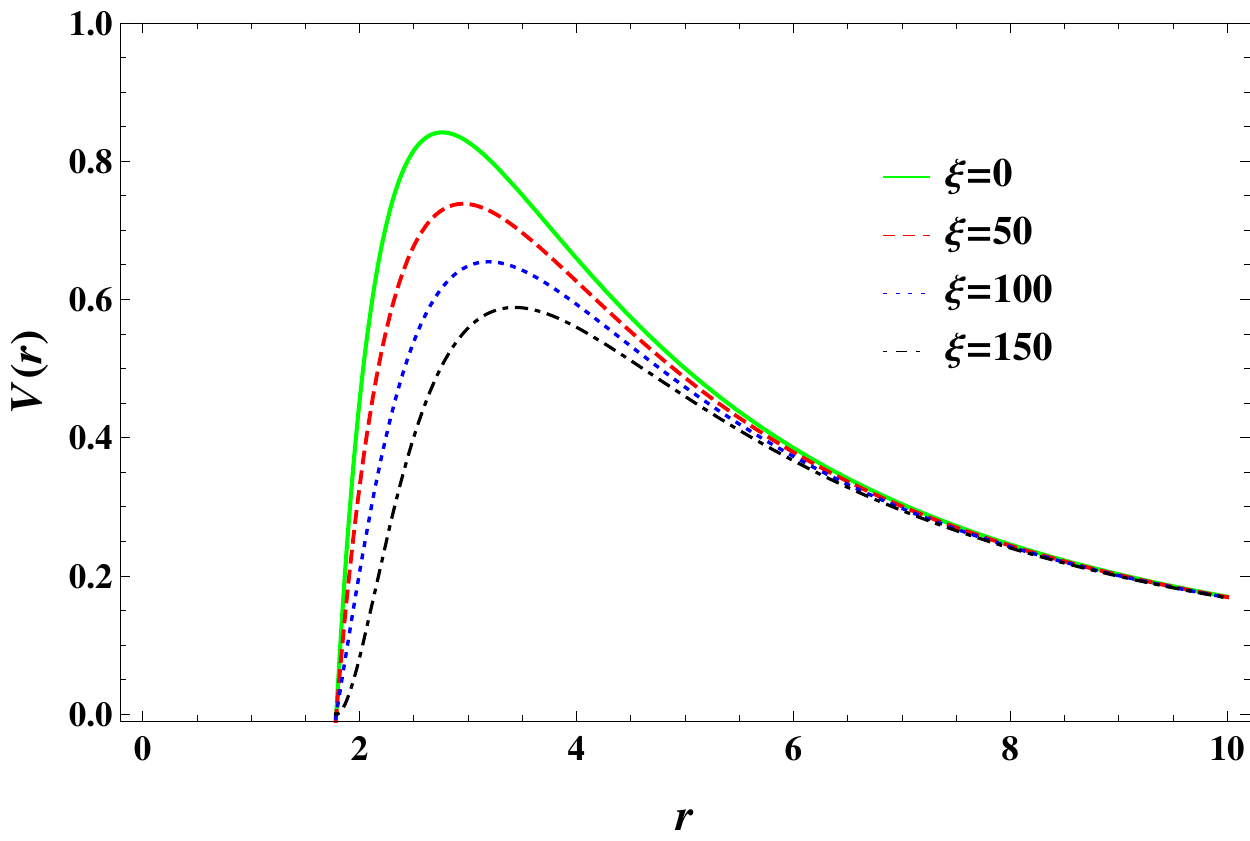}}
\subfigure[\scriptsize{Hayward}]
{\includegraphics[width=.49\textwidth]{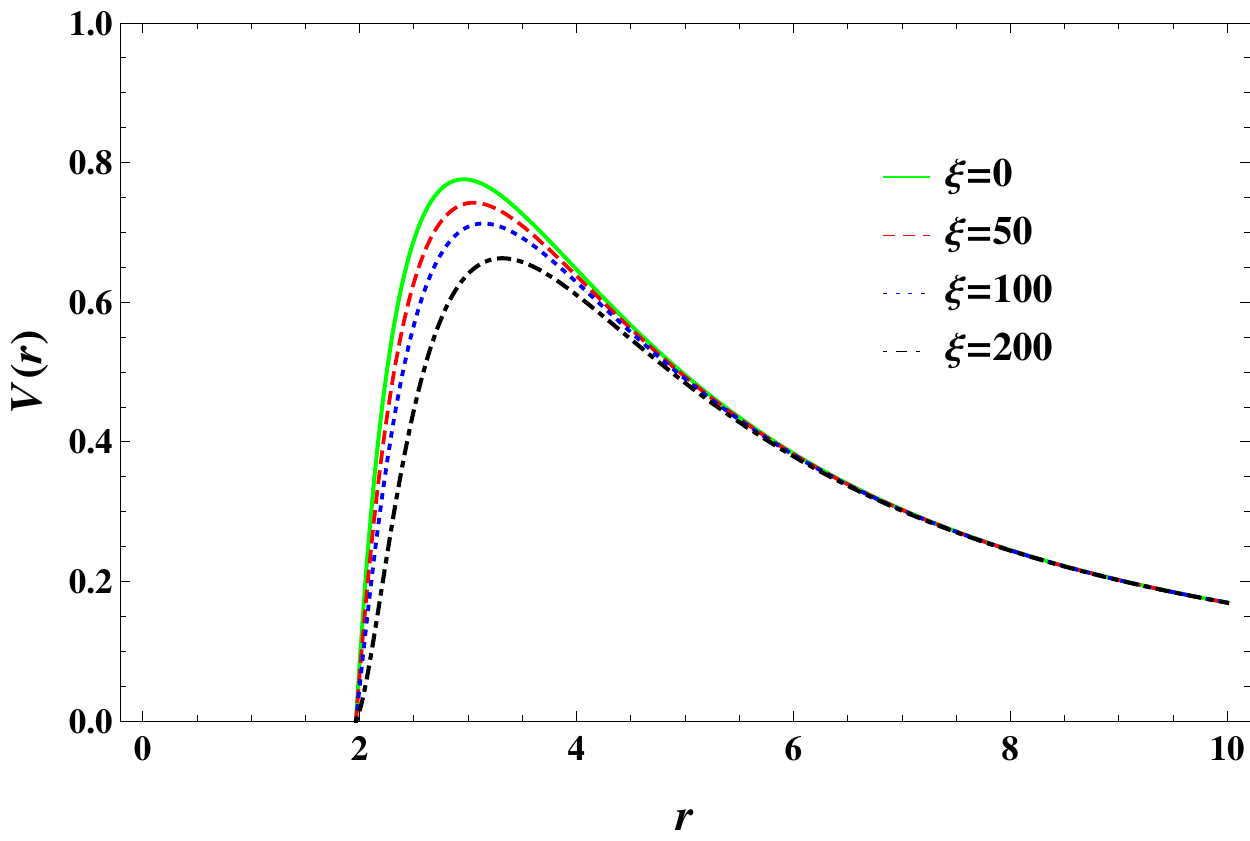}}
\subfigure[\scriptsize{ABG}]
{\includegraphics[width=.49\textwidth]{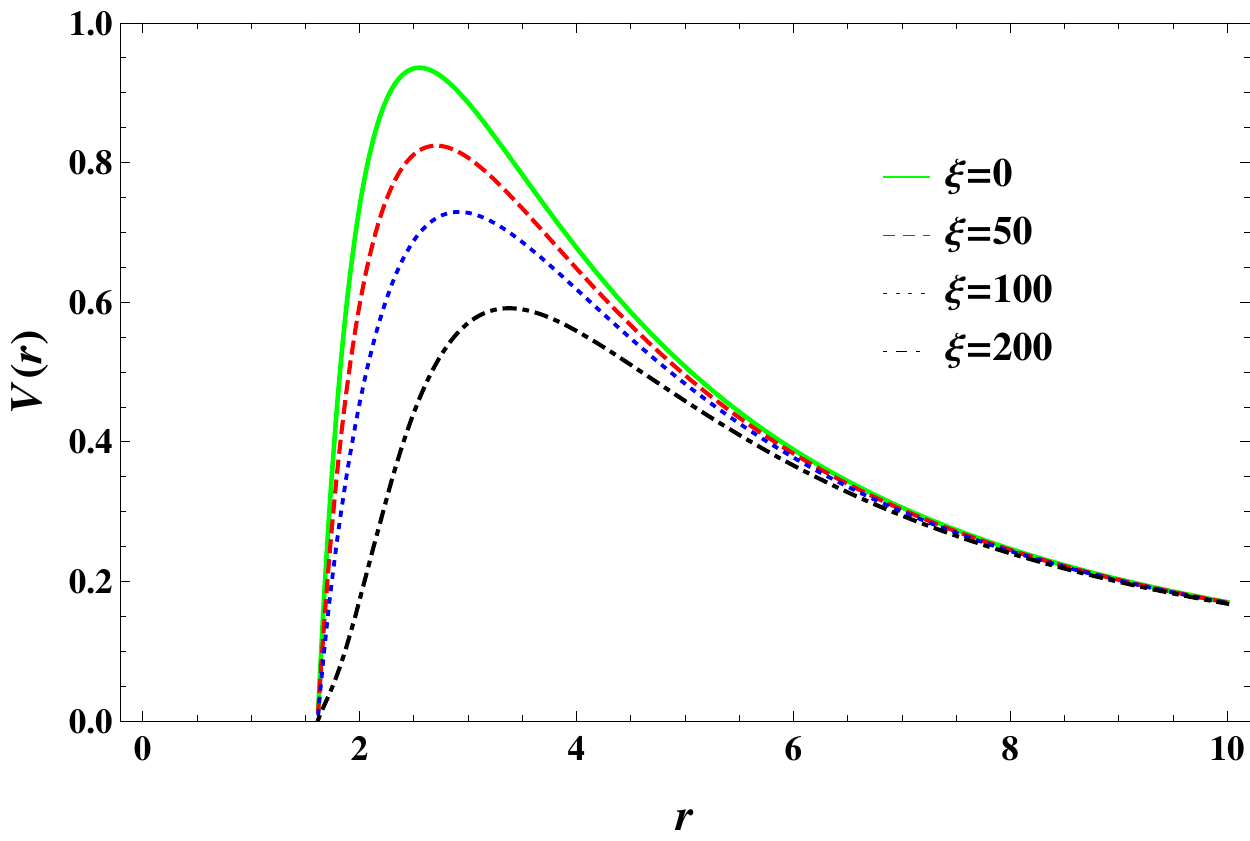}}
		    \caption{\footnotesize{ Effective potential for different coupling constants $\xi$ when $M=1$, $\mu=0.1$, $q=0.5$, and $l=4$.}}
\label{f2}
\end{figure}

\begin{figure}[H]
\centering
\subfigure[\scriptsize{Bardeen}]
{\includegraphics[width=.32\textwidth]{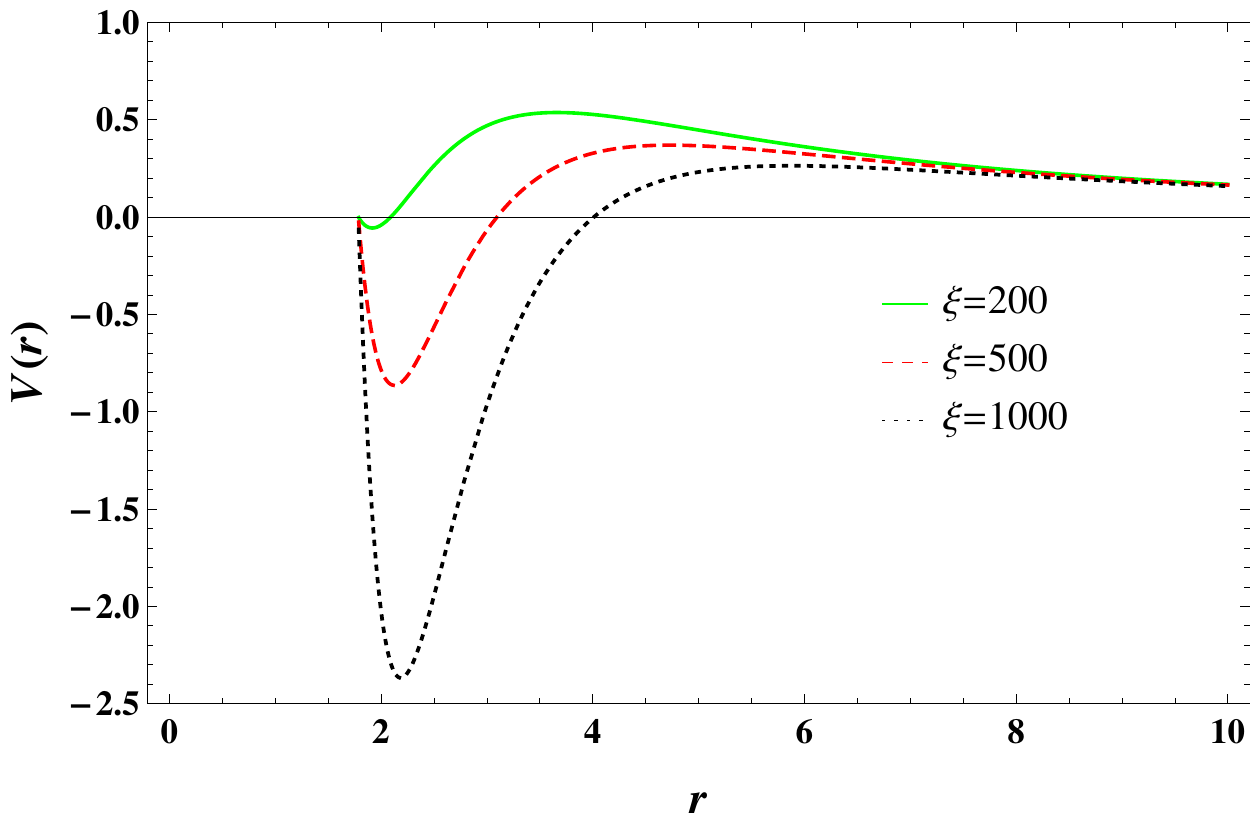}}
\subfigure[\scriptsize{Hayward}]
{\includegraphics[width=.33\textwidth]{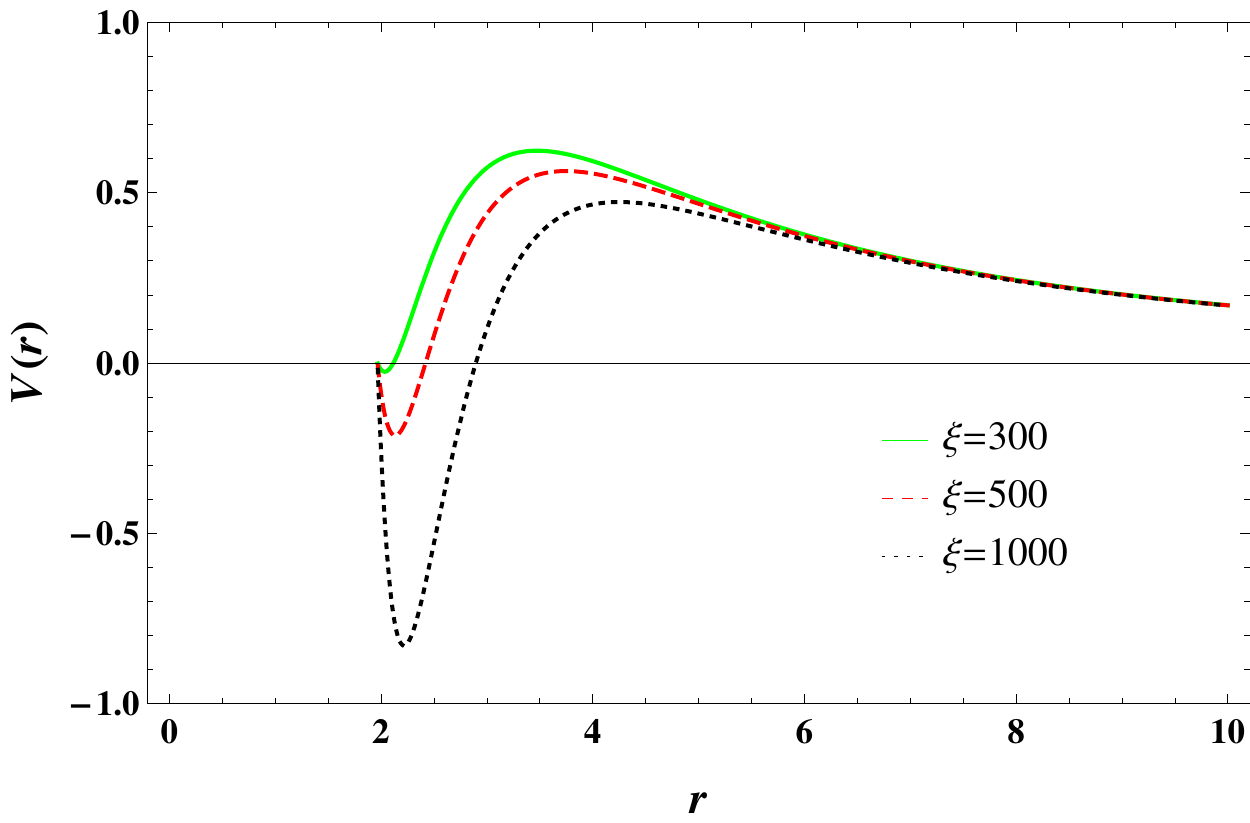}}
\subfigure[\scriptsize{ABG}]
{\includegraphics[width=.33\textwidth]{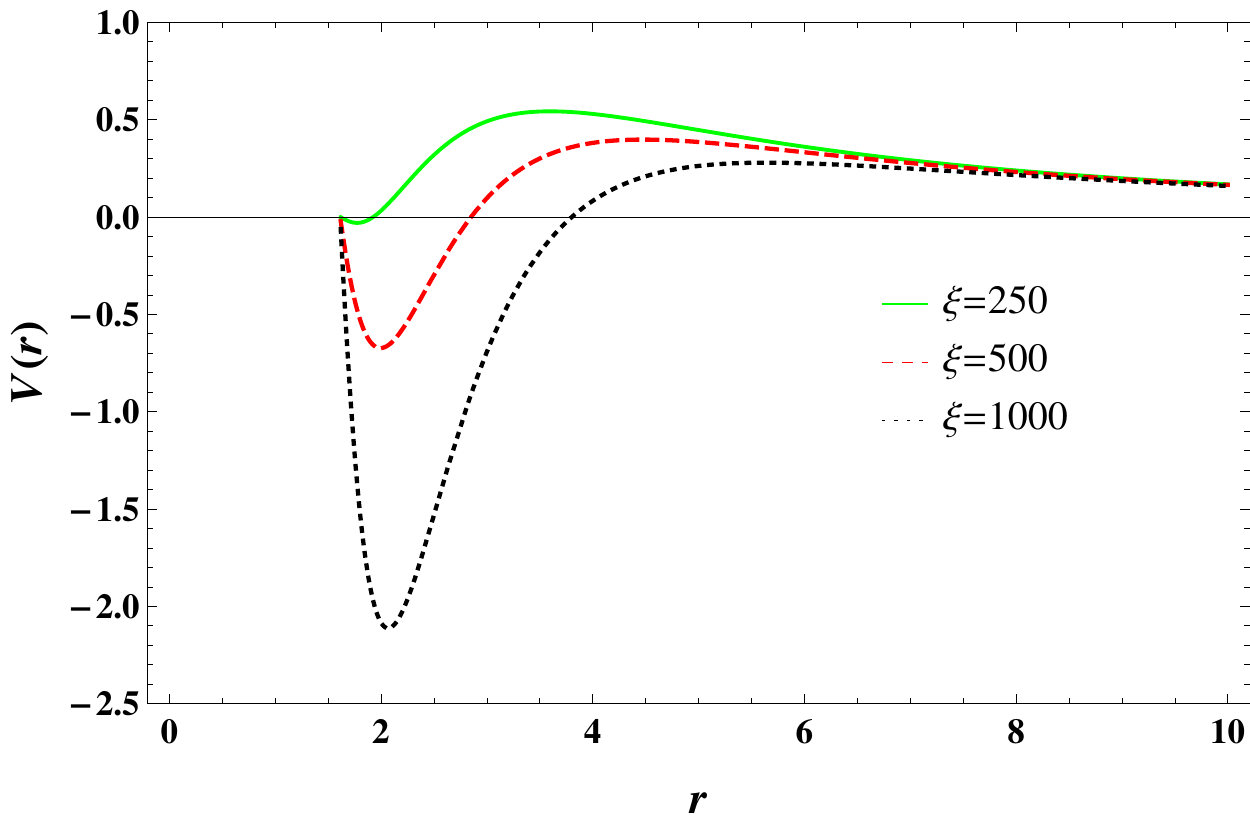}}
	    \caption{\footnotesize{ The behavior of effective potential at large $\xi$. The larger value of $\xi$ corresponds to deeper well.}}
\label{f3}
\end{figure}

\begin{figure}[H]
\centering
\subfigure[\scriptsize{Bardeen}]
{\includegraphics[width=.45\textwidth]{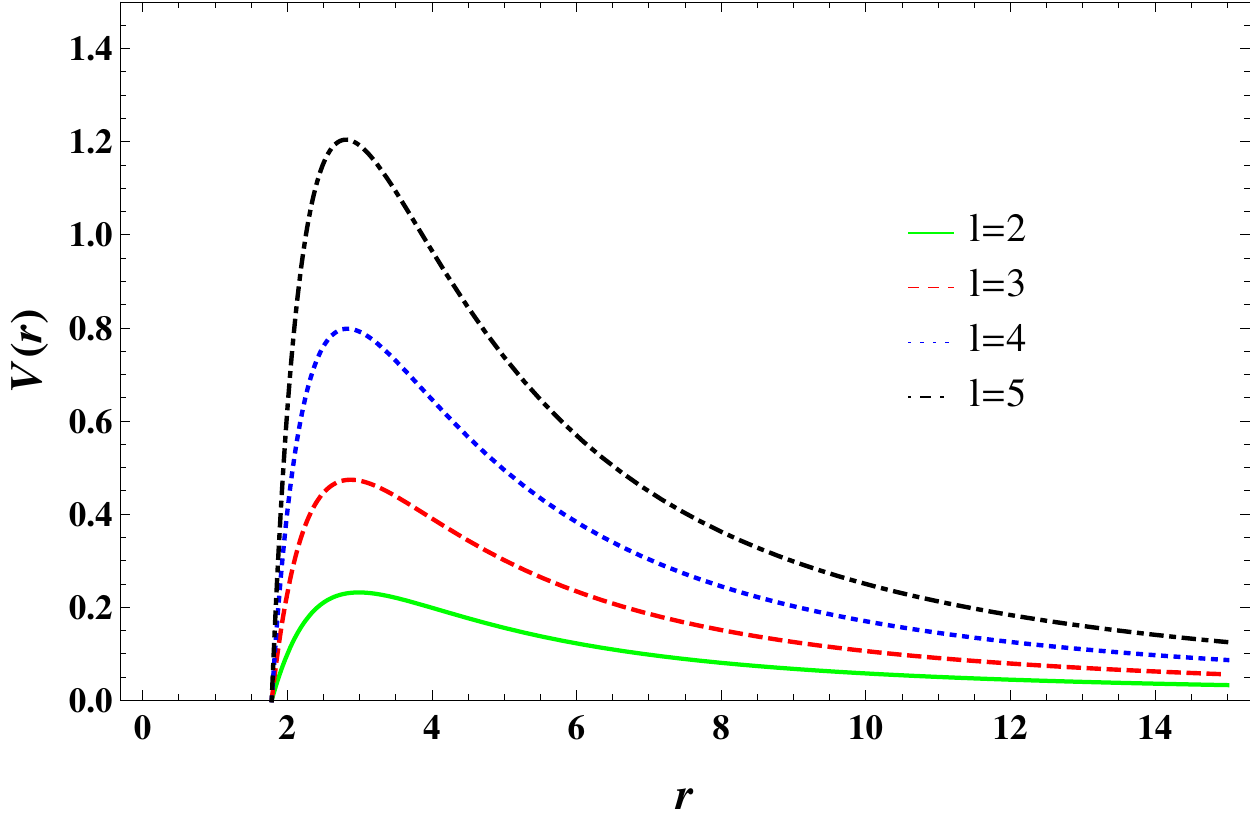}}
\subfigure[\scriptsize{Hayward}]
{\includegraphics[width=.45\textwidth]{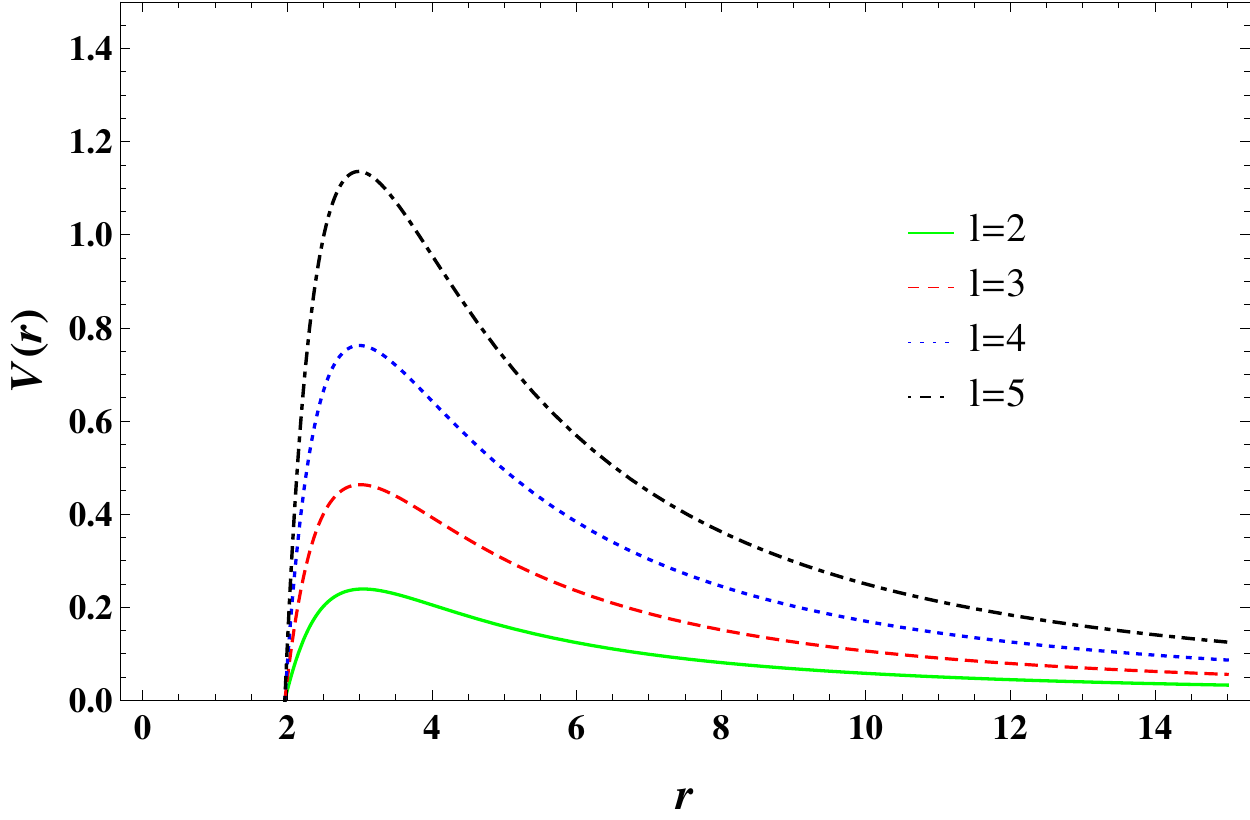}}
\subfigure[\scriptsize{ABG}]
{\includegraphics[width=.45\textwidth]{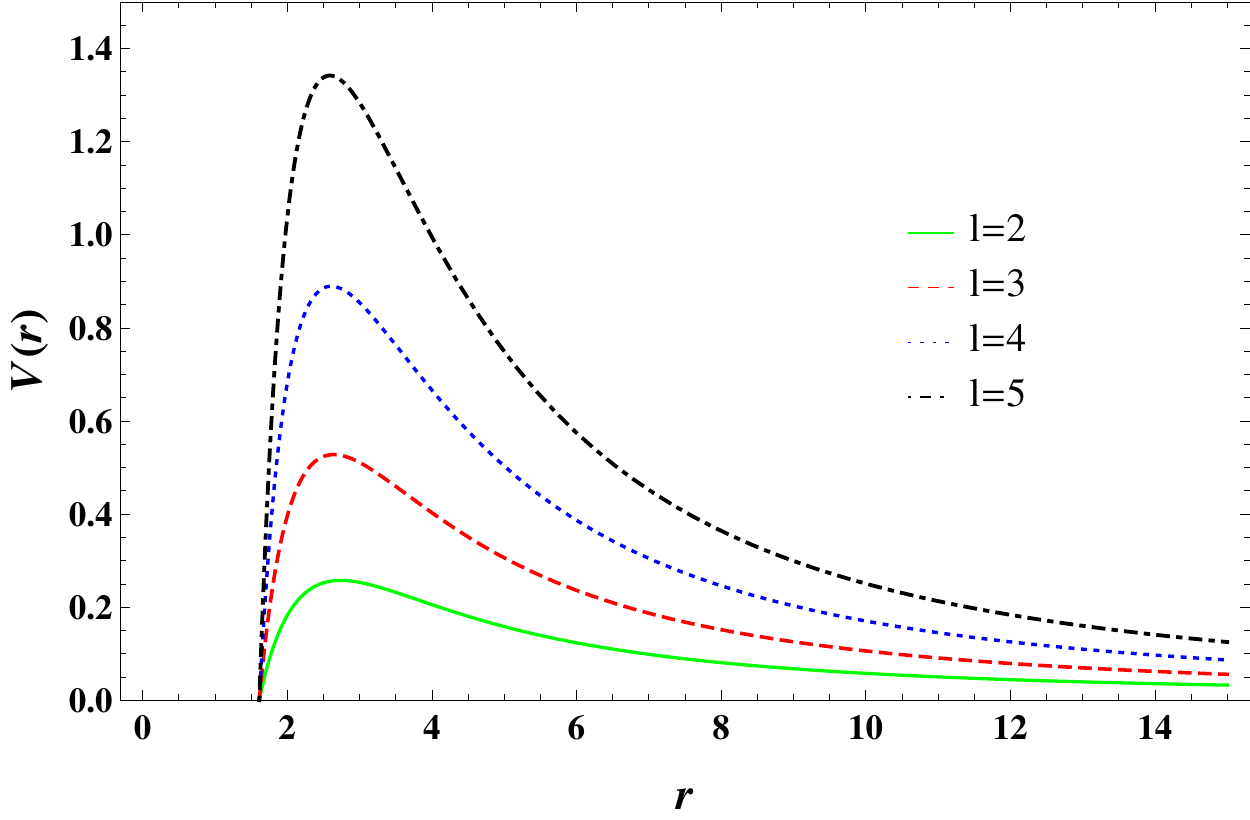}}
	    \caption{\footnotesize{ Effective potential for different multipole number $l$ when $M=1$, $\mu=0.1$, $q=0.5$, and $\xi=20$.}}
\label{f4}
\end{figure}

\begin{figure}[H]
\centering
\subfigure[\scriptsize{Bardeen}]
{\includegraphics[width=.45\textwidth]{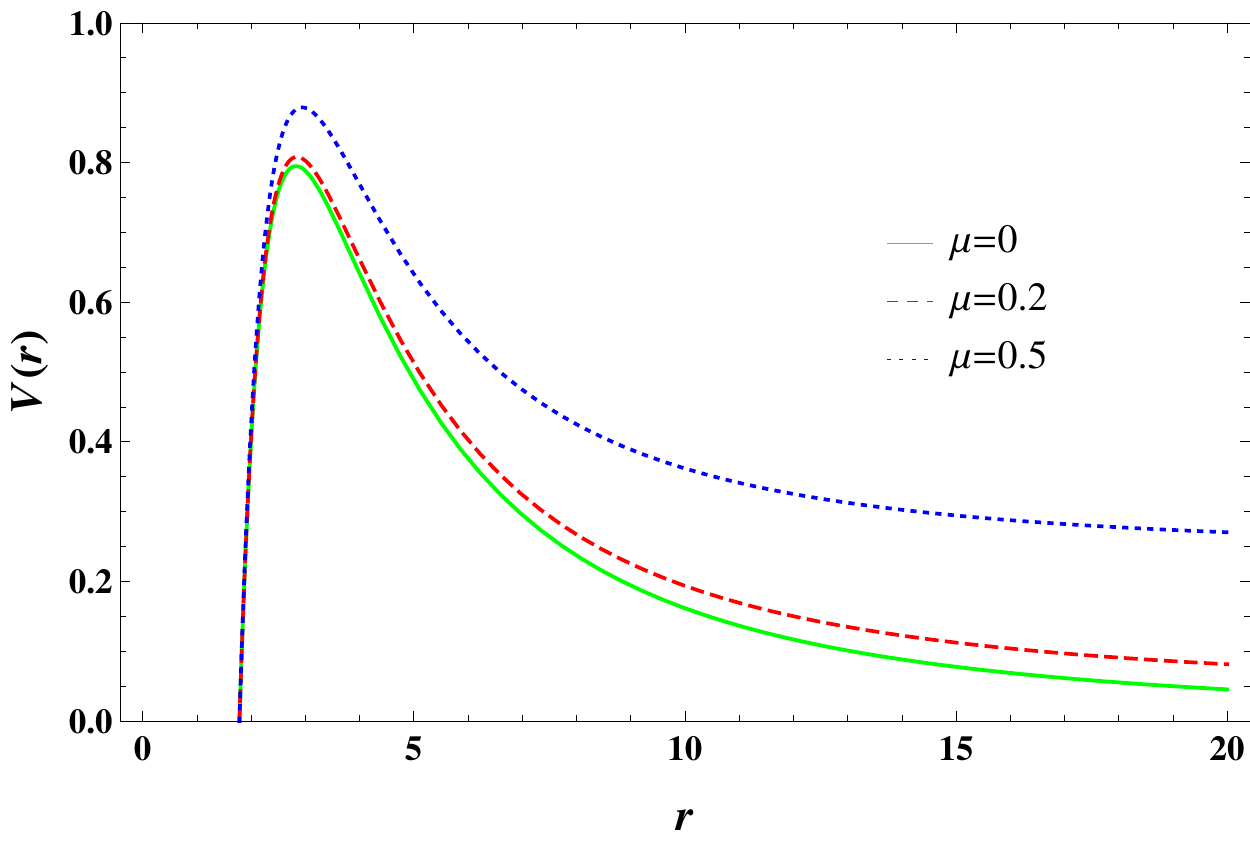}}
\subfigure[\scriptsize{Hayward}]
{\includegraphics[width=.45\textwidth]{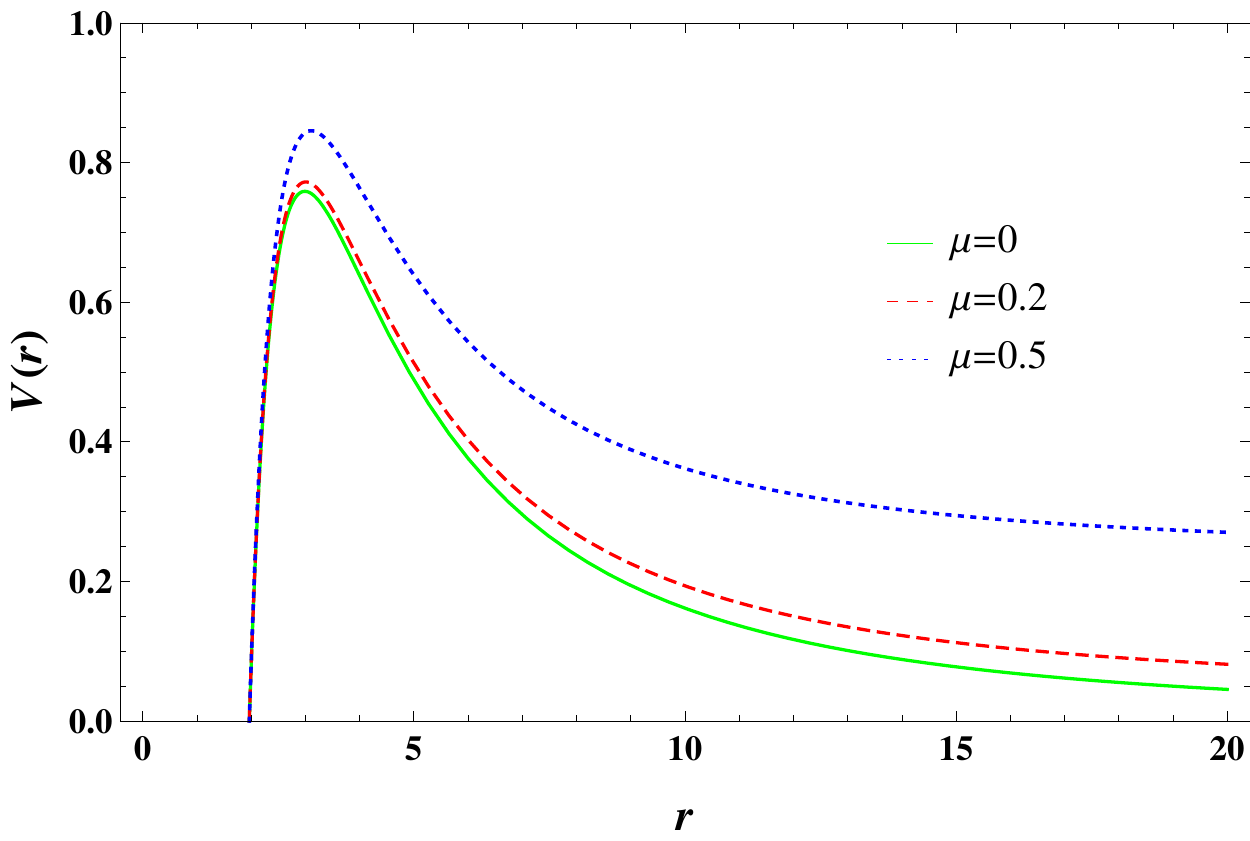}}
\subfigure[\scriptsize{ABG}]
{\includegraphics[width=.45\textwidth]{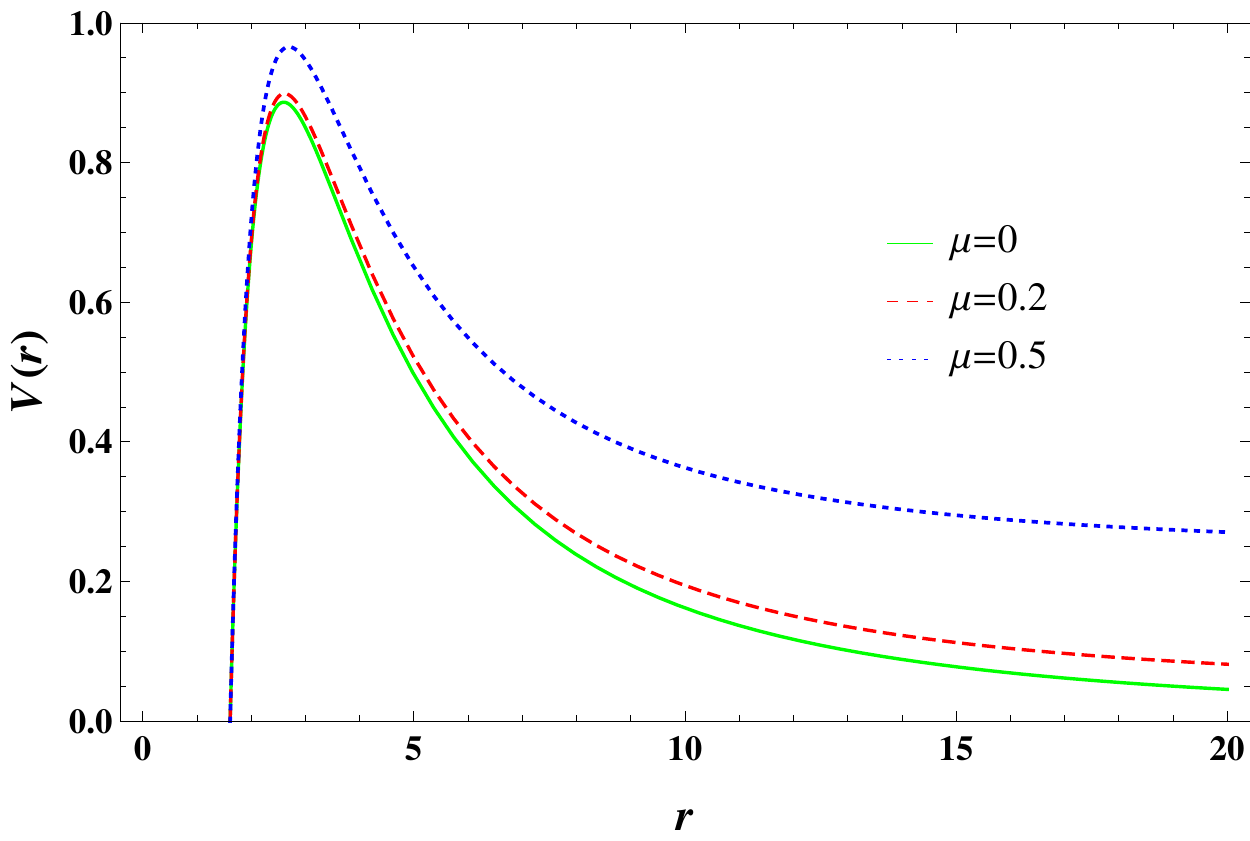}}
	    \caption{\footnotesize{Effective potential for different scalar masses $\mu$ when $M=1$, $l=4$, $q=0.5$, and $\xi=20$.}}
\label{f5}
\end{figure}

It has been proved in\cite{Horowitz:1999jd,Cardoso:2003pj} that the imaginary part of the frequency $\omega$ is negative, for the waves satisfying these boundary conditions which provided that the effective potential $V$ is positive, that is, if $V$ is positive definite then we should necessarily have $\omega_{I} < 0$. So, for a neutral scalar field, if the potential is positive outside the horizon then the black hole is stable under scalar field perturbation.
Note also that the effective potential (\ref{veff}) for RBHs vanishes at the event horizon where the curves coincide with each other, and goes to zero at infinity except for massive scalar fields which can be easily seen by plotting the effective potential as a function of $r$ for some values of parameters. We have illustrated the results in Figs.~(\ref{f2})-(\ref{f5}). As is obvious from the figures the potential curves have a local barrier exterior to the event horizon, so we can employ WKB approach to determine the QNM frequencies.

In Figs.~(\ref{f2}) we have considered the behavior of $V(r)$ in terms of different values of the coupling constant $\xi$. Due to the lapse function $f(r)$, the potential peaks of RBHs are essentially different For different values of $\xi$, however the larger the value of the coupling, the smaller the value of the potential peak. Comparing the plots shows that changing the coupling causes more drastic changes in ABG than other two black holes.

We have also observed that for large values of the coupling, nearly for $\xi>200$ as depicted in Figs.~(\ref{f3}), the potential has a well just outside the horizon for each RBH that might lead to superradiant instability \cite{Brito:2015oca,Witek:2012tr,Gonzalez:2017shu} and WKB method cannot be used as a good approximation to find the QNMs spectra. Because of the negative potential we have some unstable modes correspond to the bound states and the analytic continuation in this case should be done in a different way \cite{Brito:2015oca}. As noticed before, the shape of the potential depends on the space parameters, so to remedy this issue, we suggest to increase $l$ such that the well be disappeared outside the horizon. In this work we do not intend to consider this issue further.

In contrast to the behavior of the coupling constant, the hight of the effective potential peak increases when the value of multipole number $l$ increases as shown in Figs.~(\ref{f4}). In this regard, the WKB approximation developed in \cite{Iyer:1986np,Iyer:1986nq} works well once $l> n$. In other words, in many cases the WKB formula did not allow one to compute $n \geq l$ modes with a satisfactory accuracy \cite{Konoplya:2003ii}.

Since the RBHs are asymptotically flat, it is found that in the asymptotic limit the effective potential exhibits the following behavior
$ V(r\rightarrow\infty)\sim\mu^2$.
It is observed from Figs.~(\ref{f5}) that, for small masses, $V (r)$ still has the form of a potential barrier. But by increasing the amount of $\mu$, the peak of the potential increases slowly enough, so that eventually the height of the peak is lower than the asymptotic value of $\mu^2$. That is, further increases of the mass turns the potential barrier into a potential step. Due to this fact there is no turning points and we can not apply the WKB approximation correctly \cite{Iyer:1986np,Iyer:1986nq}.

In summary, all of the above potentials given in Figs.~ (\ref{f2}), (\ref{f4}) and (\ref{f5}) are real and positive outside the event horizon. Hence, following the arguments by Chandrasekhar \cite{chandra} the RBHs can be considered stable classically under perturbations by a neutral massive scalar field and we can employ the WKB method to extract the QNMs spectra.
 \section{QNMs of RBHs from WKB approximation}
The WKB approximation is a promising technique for determining the QNM frequencies semi-analytically. It can be used for solving the scattering problem, which is necessary to find GFs of the black hole, and for the calculation of QNMs \cite{Konoplya:2019hlu}. The main motivation for using this method is the similarity between the equation of perturbation theory for a particle and the one-dimensional Schr\"{o}dinger equation for a potential barrier, for instance the equation (\ref{sceq}) that we obtained in the previous section. The details of this method and matching the boundary conditions to obtain QNMs are completely explained in Refs. \cite{Schutz:1985zz,Iyer:1986np}, and developed to higher orders in Refs.~\cite{Konoplya:2003ii,Konoplya:2019hlu}.
We employ the 3th and 6th order WKB formula to calculate the QNM frequencies of the RBH solutions. In order to investigate the better order of approximation we will also consider the error estimation for each frequency of particular mode $n$, obtained with the WKB formula of the order $k$, especially for different values of $l$ and $\xi$.

Following \cite{Konoplya:2003ii},
 the QNM frequencies can be evaluated with the following formula
\be\label{qnm}
\frac{i\,(\omega^2-V_{\circ})}{(-2V''_{\circ})^{1/2}}-\sum_{i=2}^{6} \Lambda_i=n+\frac12,
\ee
where $\Lambda_i$ are constant coefficients resulting from higher order WKB corrections and $n=0,1,2,\dots$ is the overtone number. These coefficients are lengthy expressions containing the value and derivatives of the effective potential $V$ evaluated at $r_\circ$, which corresponds to the location of the potential peak. The explicit expressions of $\Lambda_i$ upto 6th order are given in Refs.~\cite{Iyer:1986np,Konoplya:2003ii} and for higher orders in Ref.~\cite{Konoplya:2019hlu,Matyjasek:2017psv}.
\subsection{Numerical analysis}
\begin{figure}[H]
\centering
\includegraphics[width=8cm,height=4.9cm]{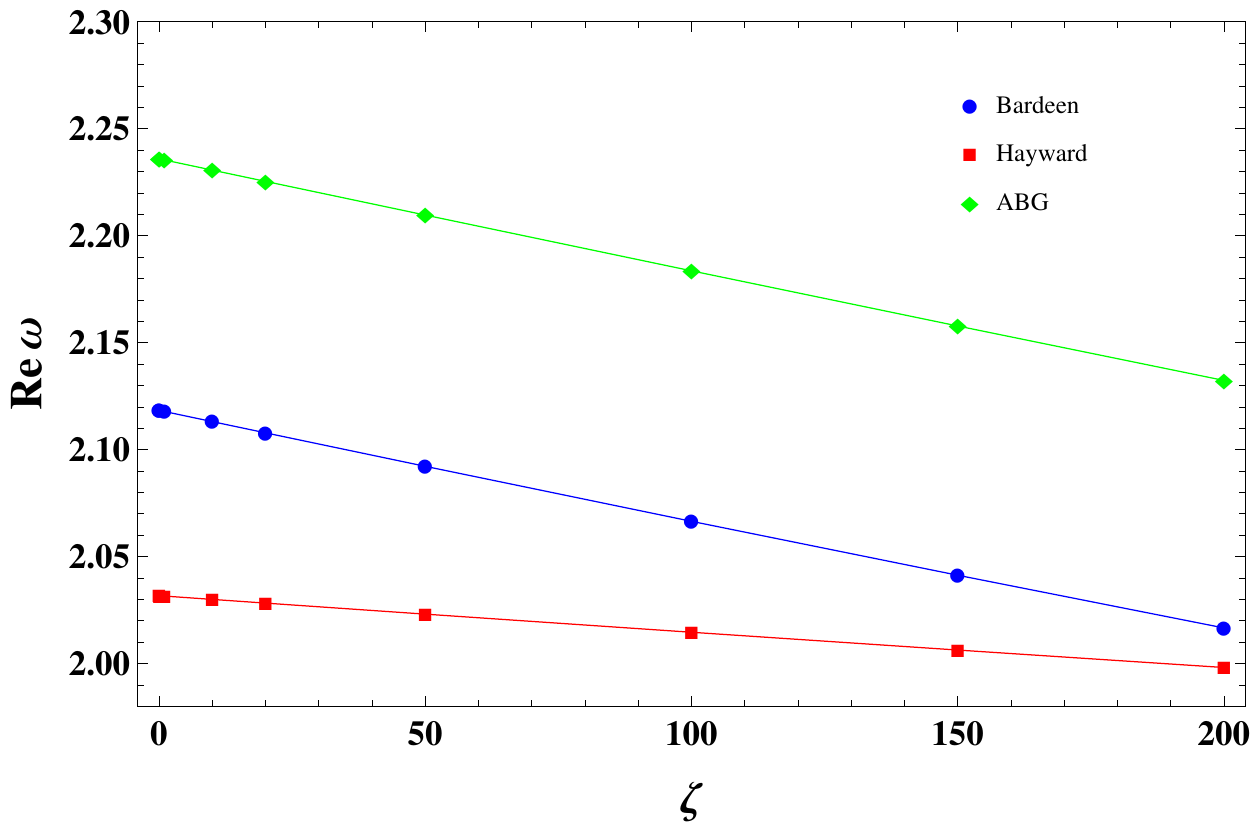}
\caption{$\omega_{R}$ vs. $\xi$ for $\mu=0.1,\,q=0.5,\,l=10,\,n=0$.}
\label{f6}
\end{figure}

\begin{figure}[H]
\centering
\subfigure[Bardeen]
{\includegraphics[width=.45\textwidth,height=4.9cm]{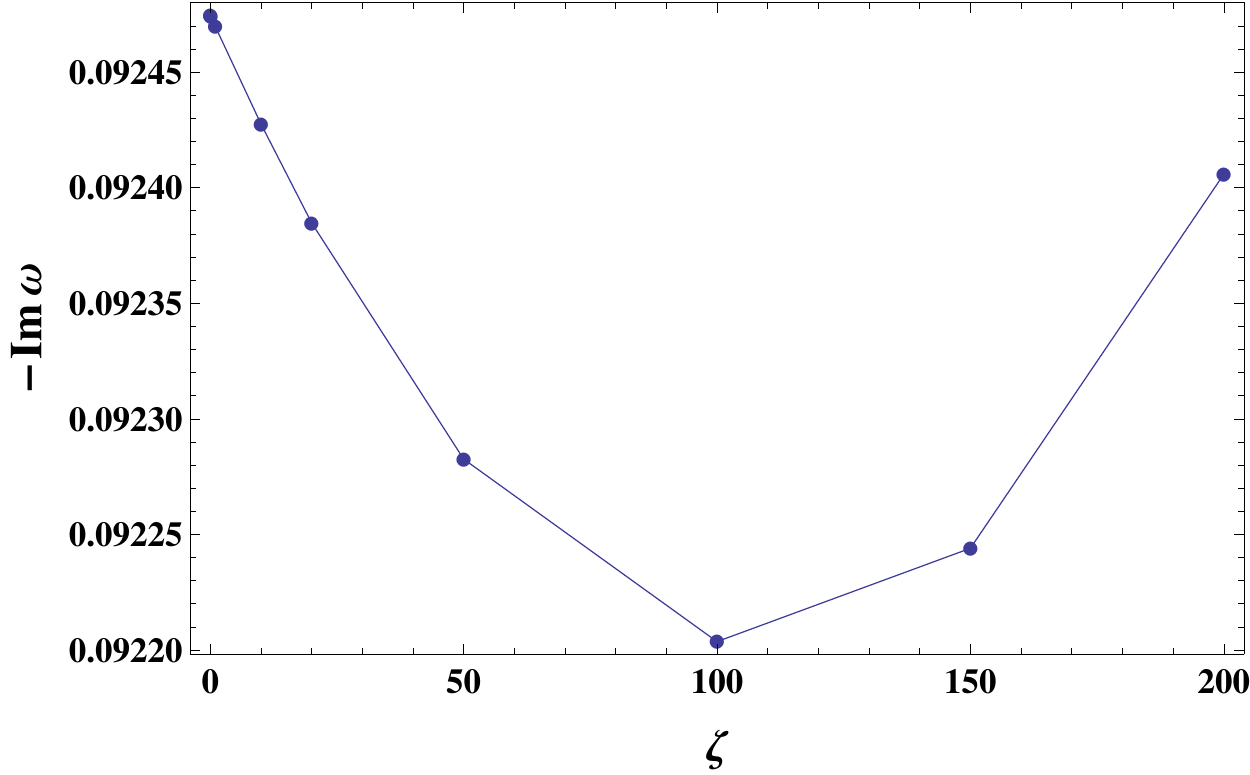}}
\subfigure[Hayward]
{\includegraphics[width=.45\textwidth,height=4.9cm]{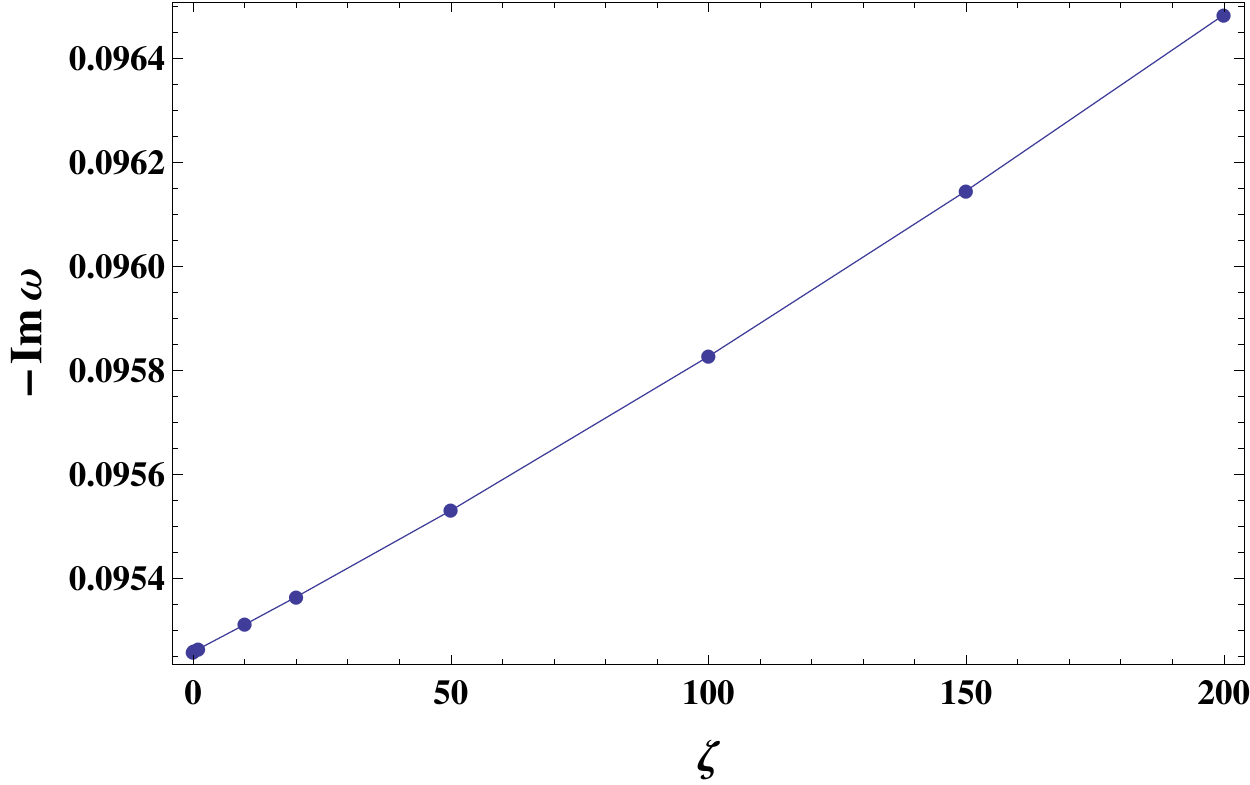}}
\subfigure[ABG]
{\includegraphics[width=.45\textwidth,height=4.9cm]{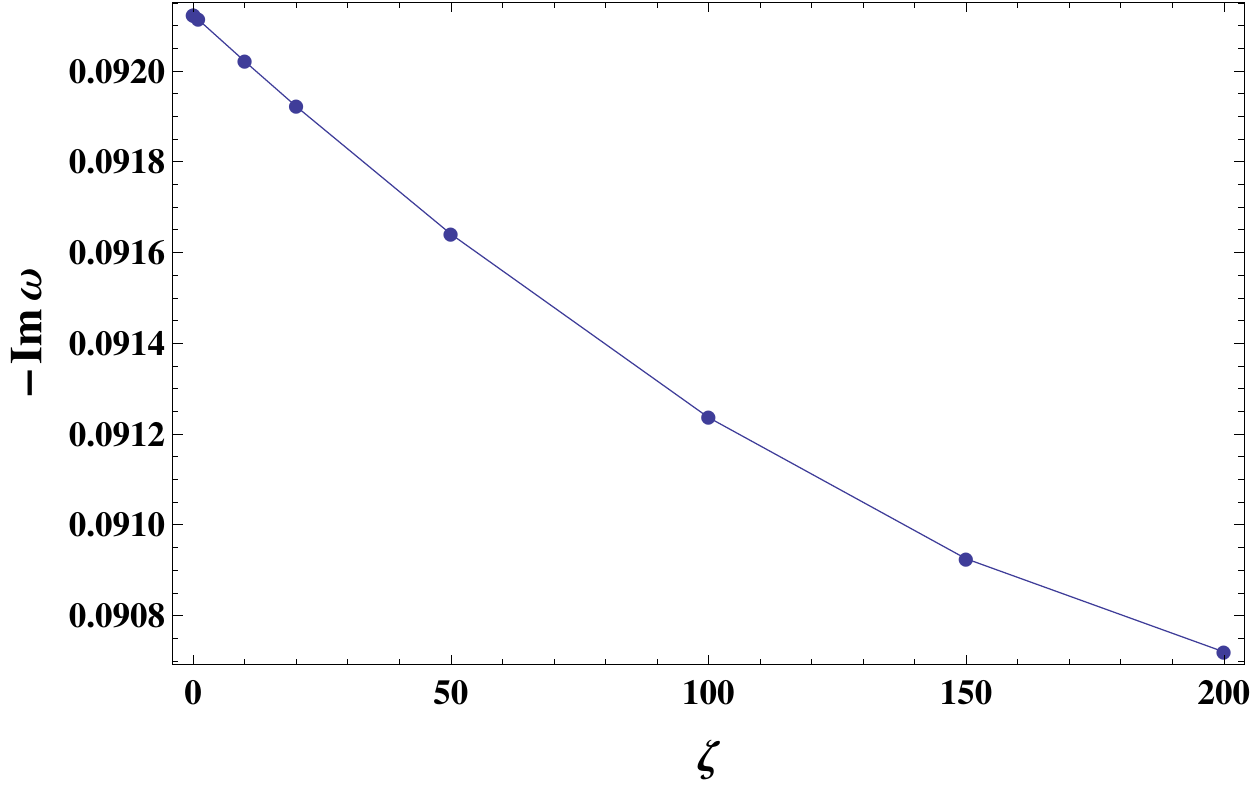}}
	    \caption{{$-\omega_{I}$ vs. $\xi$ for $\mu=0.1,\,q=0.5,\,l=10,\,n=0$.}}
\label{f7}
\end{figure}

\begin{figure}[H]
\centering
\subfigure[]
{\includegraphics[width=.45\textwidth,height=4.9cm]{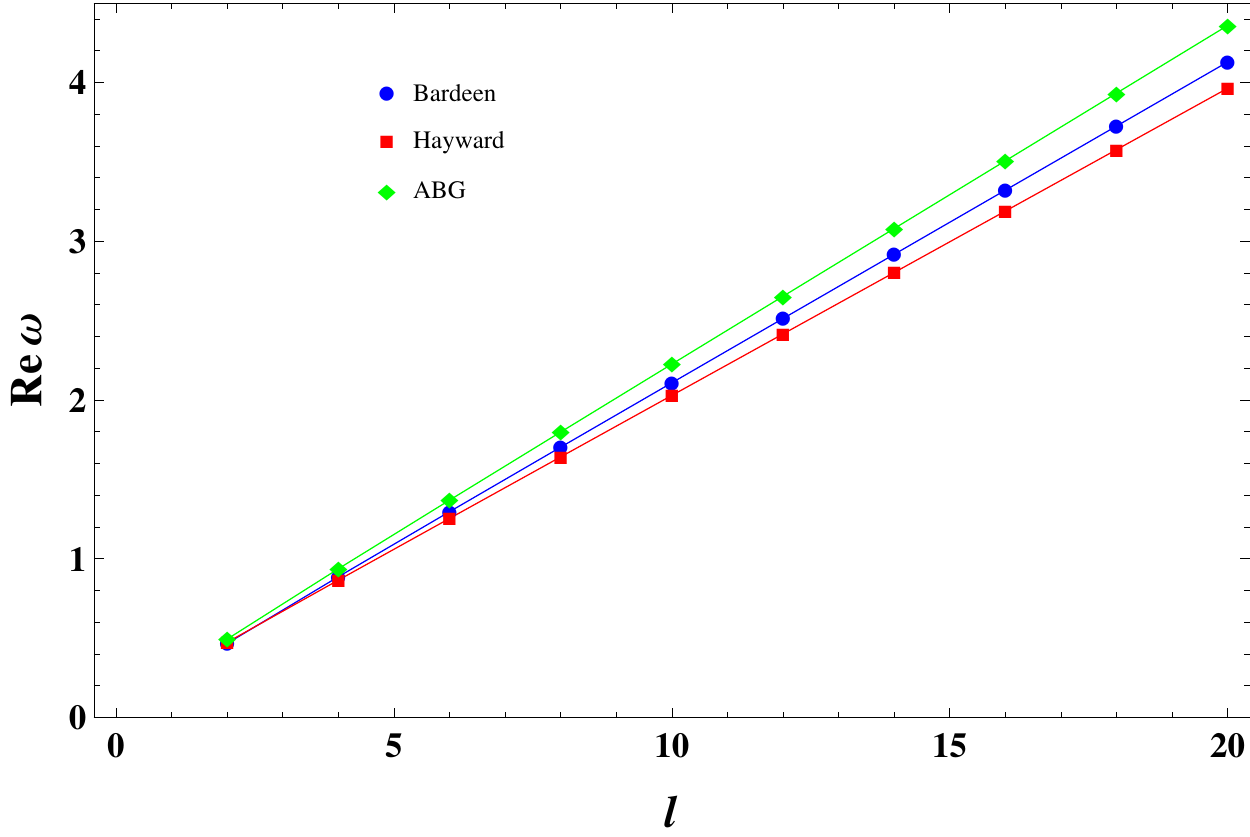}}
\subfigure[]
{\includegraphics[width=.45\textwidth,height=4.9cm]{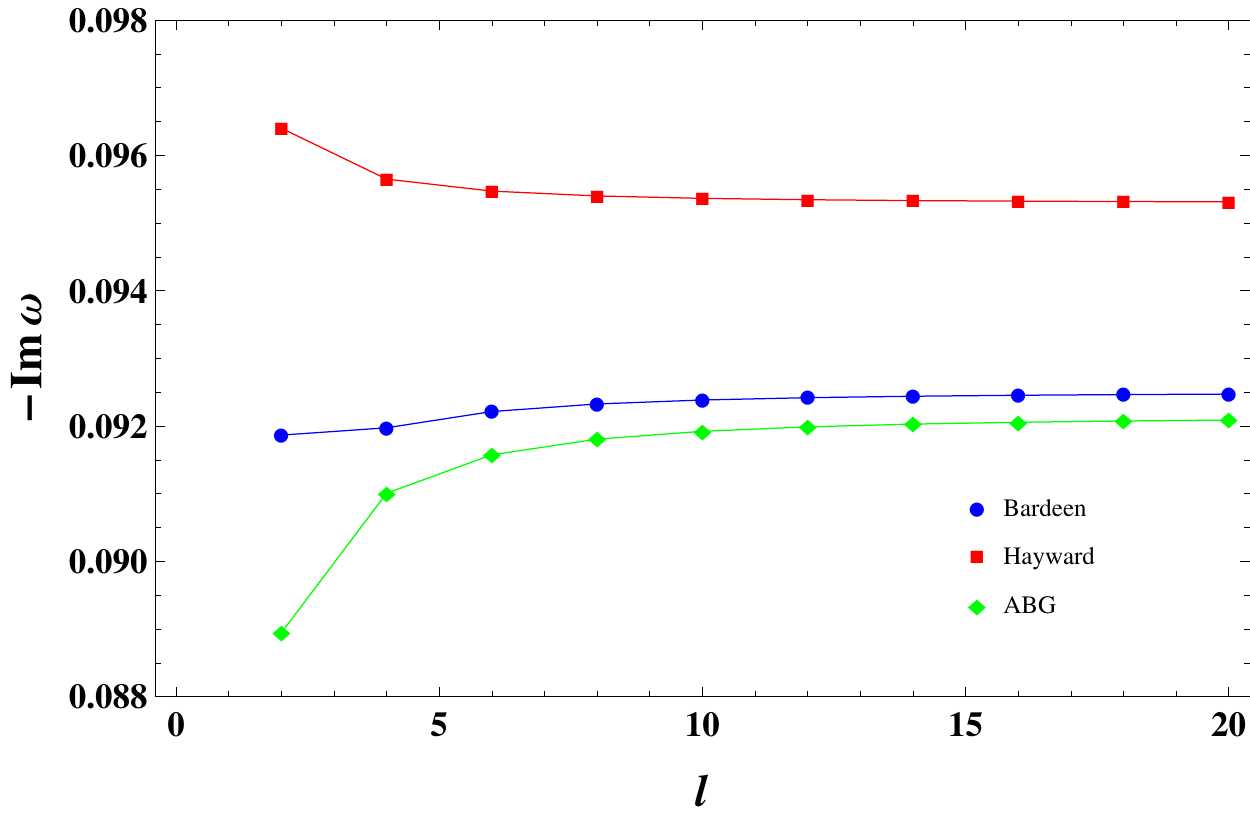}}
	    \caption{{$\omega_{R}$ and $-\omega_{I}$ vs. $l$ for $\xi=20,\, \mu=0.1,\,q=0.5,\,n=0$.}}
\label{f8}
\end{figure}

\begin{figure}[H]
\centering
\subfigure[Bardeen]
{\includegraphics[width=.49\textwidth,height=4.5cm]{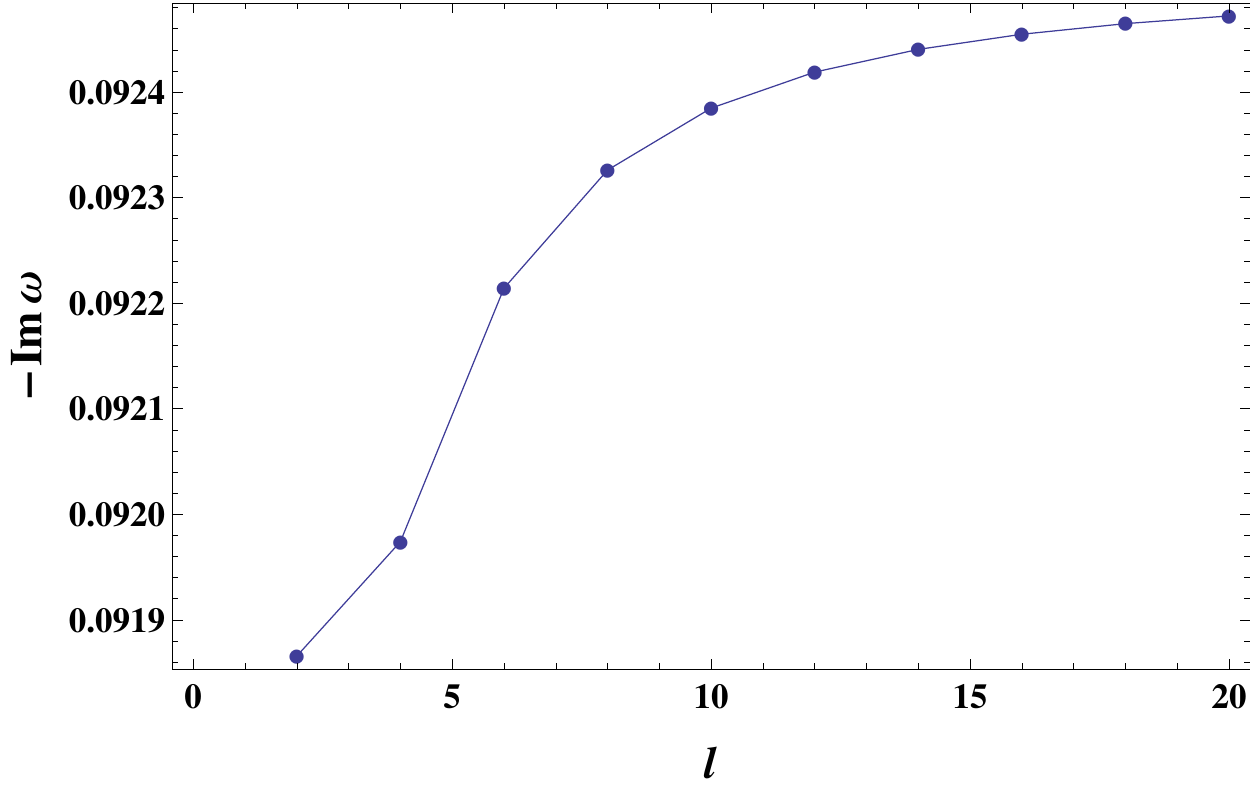}}
\subfigure[Hayward]
{\includegraphics[width=.49\textwidth,height=4.5cm]{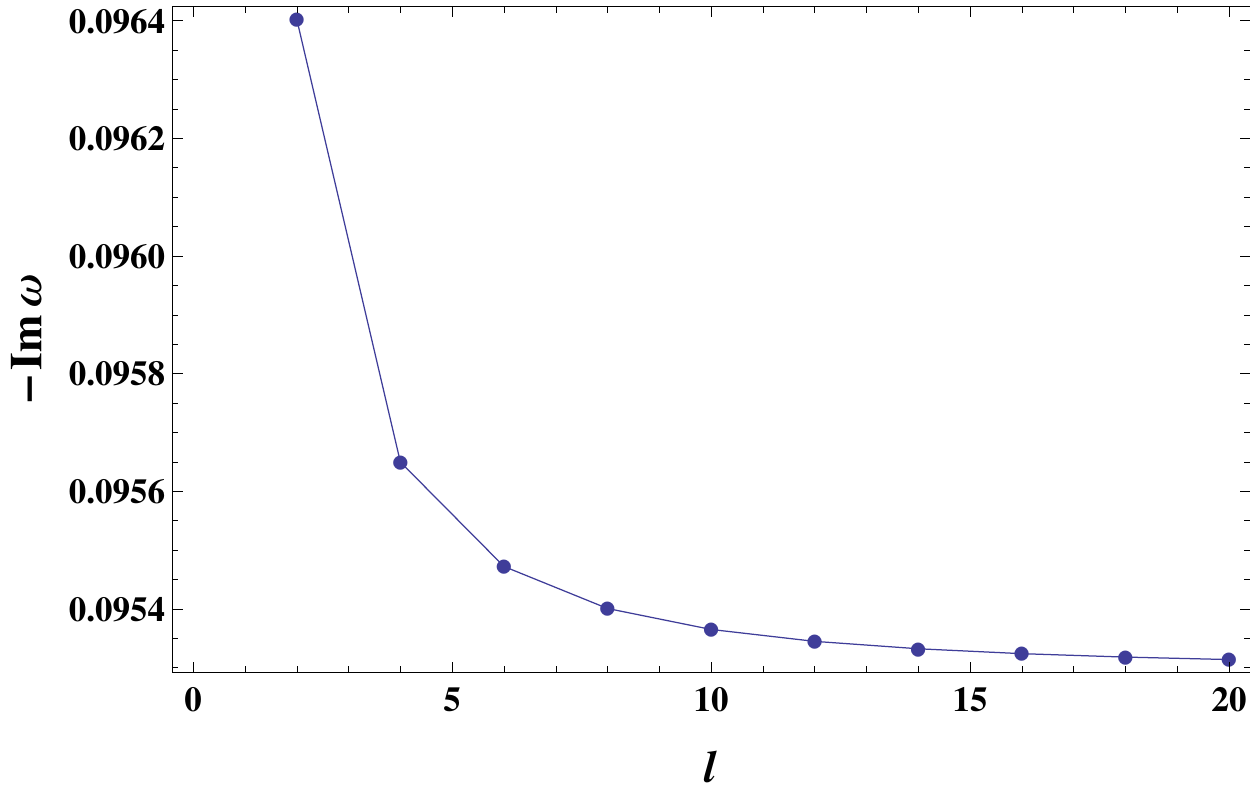}}
\subfigure[ABG]
{\includegraphics[width=.49\textwidth,height=4.5cm]{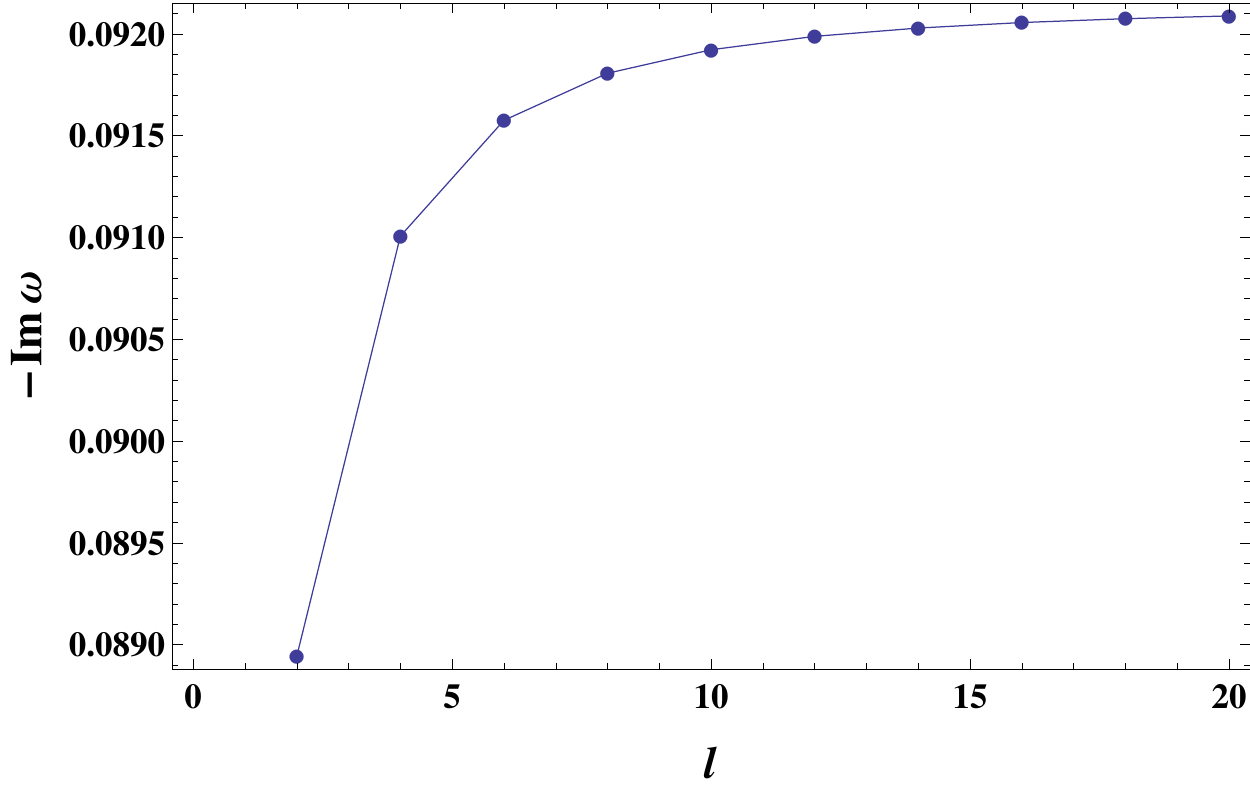}}
	    \caption{{$-\omega_{I}$ vs. $l$ for $\xi=20,\, \mu=0.1,\,q=0.5,\,n=0$.}}
\label{f9}
\end{figure}

Since the relation (\ref{qnm}) contains nontrivial functions of physical parameters, we have computed the QNM frequencies for different values of these parameters. But due to the fact that the WKB formula gives the best accuracy for $l>n$ \cite{Konoplya:2018qov}, we only consider the scalar field functions obeying this condition which are related to the low-lying QNMs. Some of the results for low-lying modes are represented for the Bardeen, Hayward, and ABG RBHs in Tabs. (\ref{tab1})-(\ref{tab4}). Without lose of generality we carry out the calculations only for $M\!=\!1$ as the RBHs masses, however one can make similar computations for different masses. In other words, all quantities in the rest of the paper are in units of the mass $M$. Since in WKB method we are unable to obtain analytical expressions for QNM frequencies, to investigate the relationship between the QNMs and physical parameters, it is common to plot these modes in terms of different values of parameters. The results are plotted in Figs.~(\ref{f6})-(\ref{f9}). It should be noted that although the figures are plotted for the 3th order approximation, the general behavior for the 6th order is the same. We have investigated that in the case of $\xi\!=\!0$ and $\mu\!=\!0$, the fundamental QNMs $(n\!=\!0)$ are exactly the values obtained in Refs. \cite{Flachi:2012nv,Fernando:2012yw} for Bardeen and Hayward black holes.

The spectrum of low-lying overtone numbers with some typical values of $\mu$ and $q$ has illustrated in Tab.~(\ref{tab1}).
As a key feature at the first glance, the imaginary parts of QNM frequencies are negative for all branches which indicates that the propagation of scalar fields in this background is stable. This property follows from the fact that the RBHs are stable due to positive potential barrier outside the event horizon.
We can see from Tab.~(\ref{tab1}) that for a particular value of $\xi$, the larger the value of $l$, the larger the value of frequencies $\omega_{R}$. However, when the coupling $\xi$ increases, $\omega_{R}$ of each overtone number $n$ decreases for a particular $l$. Though the real parts of frequencies in different regimes have the same trends for RBHs, but this is not the case for the imaginary parts. This fact can be understood from Figs.~(\ref{f6}) and (\ref{f7}).

 \begin{table}[H] 
 \caption[]{Low-lying QNM frequencies for $\mu=0.1$ and $q=0.5$. For each value of $n$, the first raw is the 3th order and the second raw is the 6th order of WKB approximation.}
\centering
\begin{tabular}{|c|c|c|c|c|c|}\hline\hline
 $l$ & $\xi$ & $n$ &  \small{Bardeen} & \small{Hayward} & \small{ABG}   \\  \hline
        & $0$     & $0$ & $0.309744 - i\, 0.091421$ & $0.296883 - i\, 0.094182$ & $0.326487 - i\, 0.091170$   \TBstrut\\
         &   &    &    $0.311600 - i\, 0.091566$ & $0.298847 - i\, 0.094202$ &  $0.328311 - i\, 0.091347$   \\ \cline{2-6}
$1$  & $20$   & $0$ &  $0.245507 - i\, 0.098074$ & $0.271175 - i\,  0.099314$ & $0.256206 -i\,0.093955 $   \TBstrut\\
         &   &    &   $0.245748 - i\, 0.099147$ & $0.270601 - i\, 0.099915$ &  $0.247202 - i\, 0.097625$    \\ \hline
       &   & $0$ & $0.509414 -i\, 0.092013$ & $0.488633 - i\,0.094748$ & $0.531411 -i\, 0.055699$   \TBstrut \\
      &  $0$ &  &    $0.509885 - i\, 0.092045$ & $0.489119 - i\, 0.094741$ &  $0.537770 - i\, 0.091769$   \\ \cline{3-6}
       &  & $1$  &  $0.491423 -i\, 0.281022$ & $0.467499- i\,0. 290326$ & $0.469899 -i\, 0.182961$ \TBstrut\\
         &   &    &   $0.492564 - i\, 0.280906$ & $0.468491 - i\, 0.290165$ &  $0.522057 - i\, 0.279275$    \\ \cline{2-6}
 2   &  & $0$  & $0.466454 -i\, 0.091866$ & $0.473445 -i\,0.096402$  & $0.493546 -i\, 0.088943$   \TBstrut \\
   &  $20$   &    &    $0.467265 - i\, 0.091642$ & $0.473939 - i\, 0.096281$ &  $0.495507 - i\,0.088757$    \\ \cline{3-6}
     & & $1$   & $0.442152 -i\, 0.281960$ & $0.448088 -i\, 0.296233$ & $0.470019 -i\, 0.271090$ \TBstrut\\
         &   &    &    $0.444293 - i\, 0.279795$ & $0.448762 - i\, 0.295556$ &  $0.480438 - i\, 0.265019$    \\ \hline
       &       & $0$ & $0.709717 -i\, 0.092242$ & $0.680751 - i\, 0.094994$  & $0.743341 -i\,0.093743 $   \TBstrut\\
         &   &    & $0.709897 - i\, 0.092251$ & $0.680933 - i\, 0.094991$ &  $0.749007 - i\, 0.091939$   \\ \cline{3-6}
       & $0$ & $1$  & $0.696987 -i\, 0.279126$ & $0.665698 - i\, 0.287925$  & $0.688595 -i\, 0.124989$\TBstrut\\
         &   &    &   $0.697442 - i\, 0.279088$ & $0.666086 - i\, 0.287872$ &  $0.737718 - i\, 0.277768$    \\ \cline{3-6}
       &       & $2$ &  $0.674841 -i\, 0.471330$ & $0.639770 - i\, 0.487178$  & $0.573579 -i\,0.235208 $ \TBstrut \\
         &   &    &   $0.674604 - i\, 0.472425$ & $0.638964 - i\, 0.488844$ &  $0.716839 - i\, 0.469026$   \\ \cline{2-6}
       &        & $0$ & $0.678662 -i\, 0.091775$ & $0.670105 -i\, 0.095862$  & $0.717613 -i\, 0.090343$   \TBstrut\\
         &   &    &   $0.678942 - i\, 0.091794$ & $0.670317 - i\, 0.095844$ &  $0.717929 - i\, 0.090428$    \\ \cline{3-6}
 $3$& $20$ & $1$ &  $0.664031 -i\, 0.277900$ & $0.653332 -i\, 0.290700$  & $0.704800 -i\, 0.272964$ \TBstrut\\
         &   &    &   $0.665025 - i\, 0.277789$ & $0.653862 - i\, 0.290522$ &  $0.706247 - i\, 0.273197$   \\ \cline{3-6}
       &        & $2$ &  $0.638314 -i\, 0.469720$ & $0.624175 -i\, 0.492227$  & $0.681710 -i\, 0.460044$\TBstrut\\
         &   &    &  $0.639501- i\, 0.470365$ & $0.623467 - i\, 0.493602$ &  $0.685311 - i\, 0.461026$     \\ \cline{2-6}
       &        & $0$ &  $0.581656 -i\, 0.100509$ & $0.634493 -i\, 0.101545$  & $0.611317 -i\, 0.097312$   \TBstrut\\
         &   &    &  $0.581732 - i\, 0.100550$ & $0.634532 - i\, 0.101578$ &  $0.611082 - i\, 0.097427$    \\ \cline{3-6}
       &$100$& $1$  & $0.562093 -i\, 0.306540$ & $0.614218 -i\, 0.309003$ & $0.589950 -i\, 0.296699$\TBstrut\\
         &   &    &  $0.561436 - i\, 0.307215$ & $0.613737 - i\, 0.309466$ &  $0.588342 - i\, 0.298110$    \\ \cline{3-6}
       &        & $2$ & $0.530748 -i\, 0.521719$ & $0.580166 -i\, 0.525156$  & $0.554309 -i\, 0.515673$ \TBstrut\\
         &   &    &    $0.526764 - i\, 0.529029$ & $0.576311 - i\, 0.530688$ &  $0.548522 - i\, 0.515222$    \\ \hline
\end{tabular}
\label{tab1}
\end{table}

From Fig.~(\ref{f7}), one can observe that RBHs have different behaviors by increasing $\xi$. In the case of Bardeen black hole, the imaginary part $-\omega_{I}$ first decreases and then increases, but for hayward and ABG black holes, it monotonically increases and decreases respectively. It can also be found from Fig.~(\ref{f6}) that for all RBHs not only the real parts decrease when the coupling increases, but also the rate of changes in ABG is more sensitive to $\xi$ than the others. In contrast, one can see from Fig.~(\ref{f8}) that for all RBHs the real part increases when one increases $l$ but the graphs for $-\omega_{I}$ again have different behaviors .

As depicted in Figs.~(\ref{f9}), by increasing the multipole number, the imaginary part of Bardeen and ABG have nearly the same behavior and increase while for Hayward black hole it decreases. But for large values of $l$, $\omega_{I}$ goes to an asymptotic value for all RBHs which is consistent with the eikonal limit \cite{Ferrari:1984zz,Ferrari:1984ozr}. In the context of AdS/CFT, the timescale for approaching to a thermal equilibrium, $\tau=1/|\omega_{I}|$, increases for Hayward while it decreases for Bardeen and ABG by increasing $l$ \cite{Horowitz:1999jd,Cardoso:2001bb,Konoplya:2002ky}.

\begin{figure}[H]
\centering
\subfigure[]
{\includegraphics[width=.48\textwidth,height=5cm]{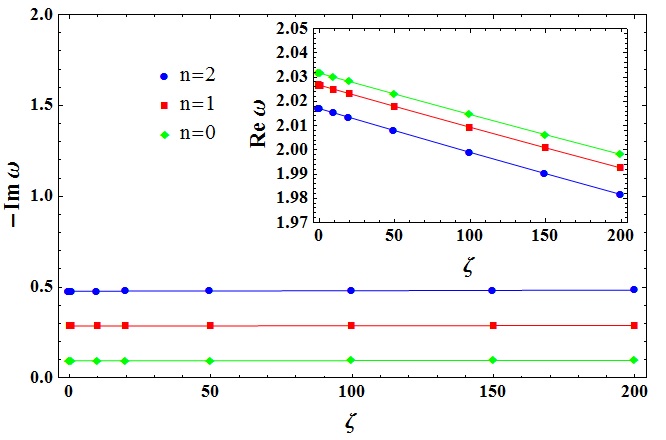}}
\subfigure[]
{\includegraphics[width=.48\textwidth,height=5cm]{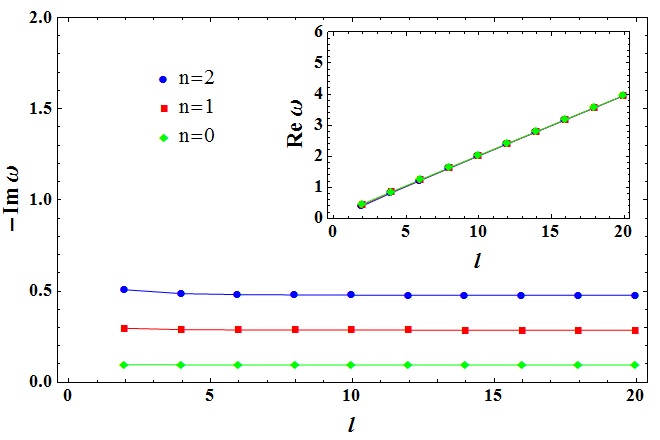}}
	    \caption{{Low-lying QNMs vs. $l$ and $\xi$ for $\mu=0.1$ and $q=0.5$.}}
\label{f10}
\end{figure}

Also the plots for low-lying modes $n=0,1,2$ in terms of these parameters are depicted in Figs.~(\ref{f10}) for Hayward black hole and we have investigated that the results are the same for the other two solutions. Comparing the plots shows that the variations of the imaginary parts are very small for different values of $\xi$ and $l$.

\begin{figure}[H]
\centering
\subfigure[]
{\includegraphics[width=.48\textwidth]{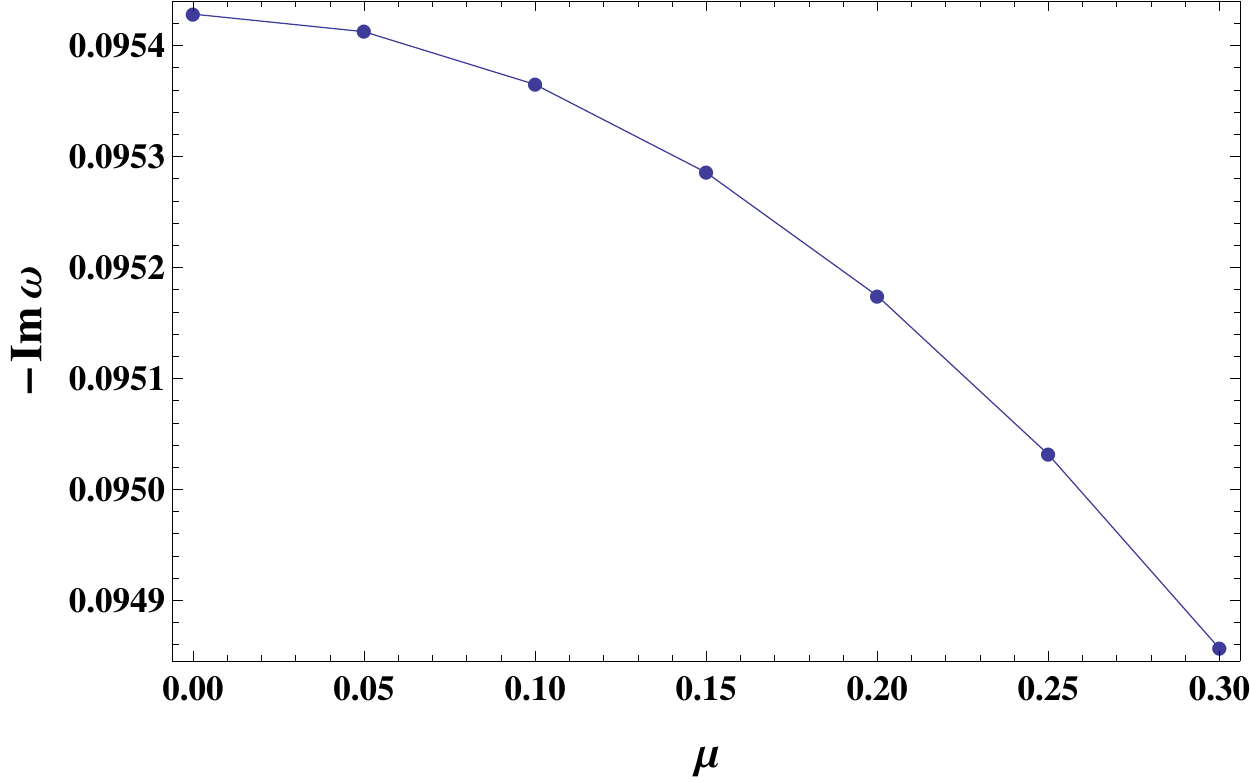}}
\subfigure[]
{\includegraphics[width=.48\textwidth]{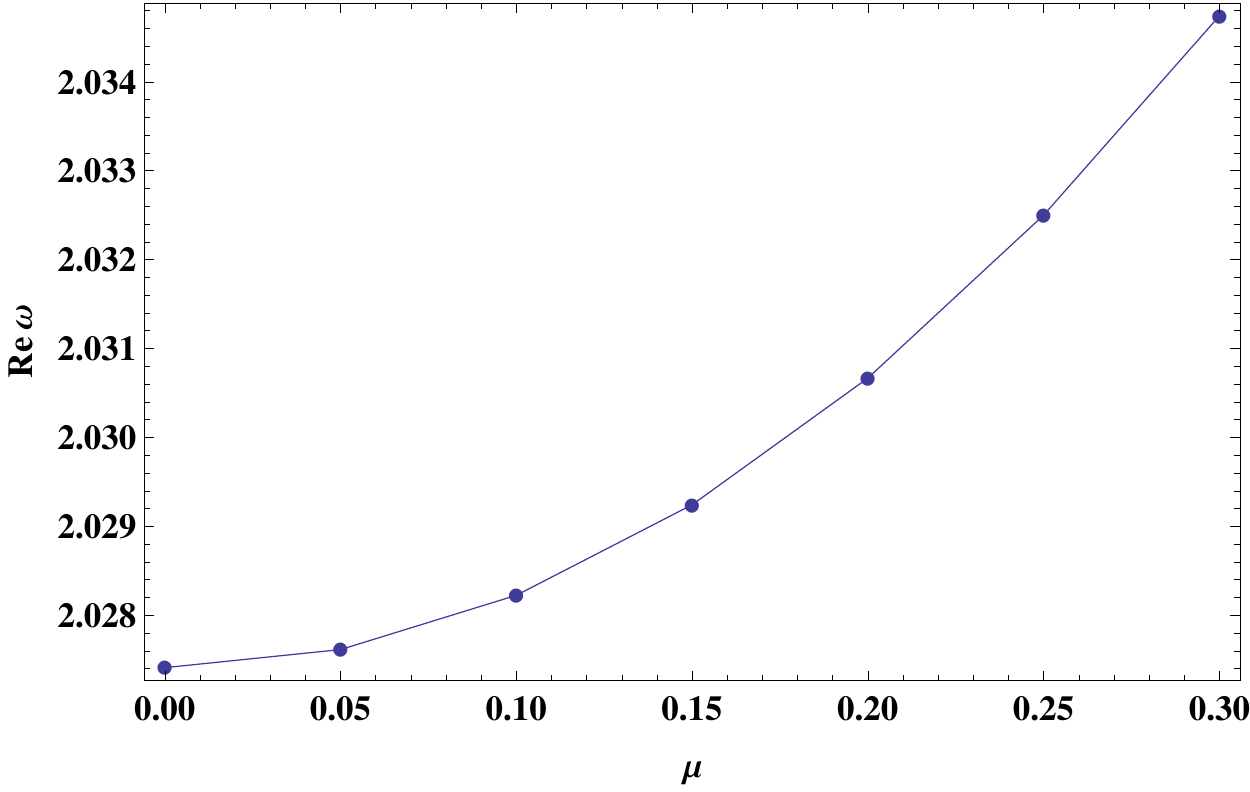}}
	    \caption{{$\omega_{R}$ and $-\omega_{I}$ in Hayward spacetime vs. $\mu$ for $\xi=20,\, l=10,\,q=0.5,\,n=0$.}}
\label{f11}
\end{figure}

\begin{figure}[H]
\centering
\subfigure[]
{\includegraphics[width=.48\textwidth]{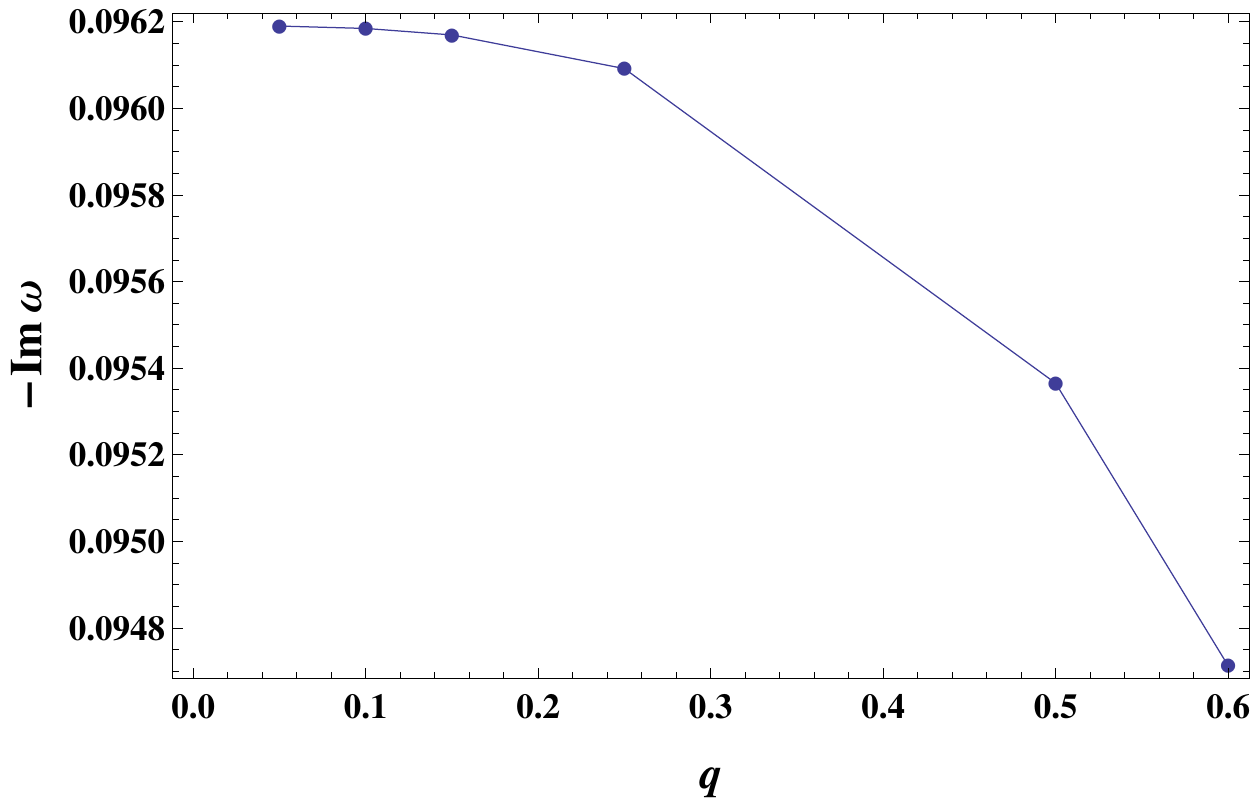}}
\subfigure[]
{\includegraphics[width=.48\textwidth]{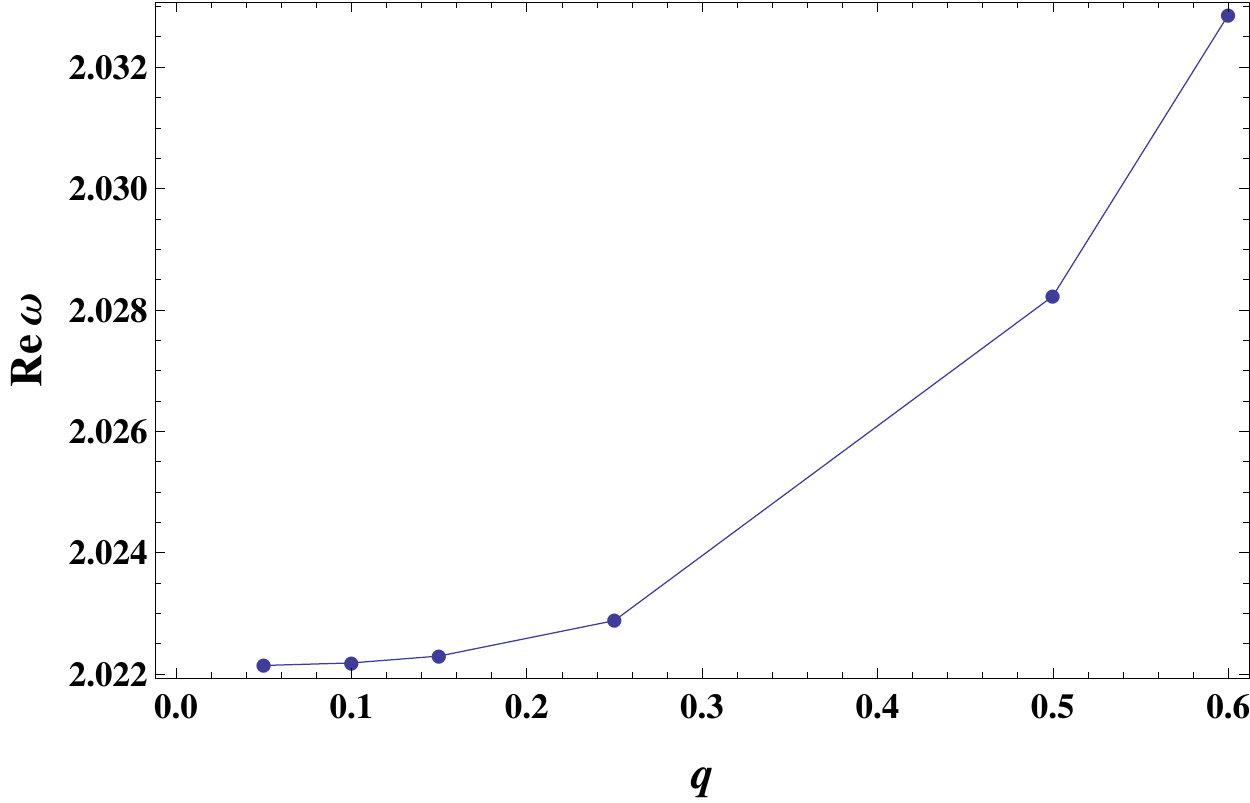}}
	    \caption{{$\omega_{R}$ and $-\omega_{I}$ in Hayward spacetime vs. $\mu$ for $\,\xi=20,\, l=10,\,\mu=0.1,\,n=0$.}}
\label{f12}
\end{figure}

\begin{figure}[H]
\centering
\subfigure[]
{\includegraphics[width=.48\textwidth]{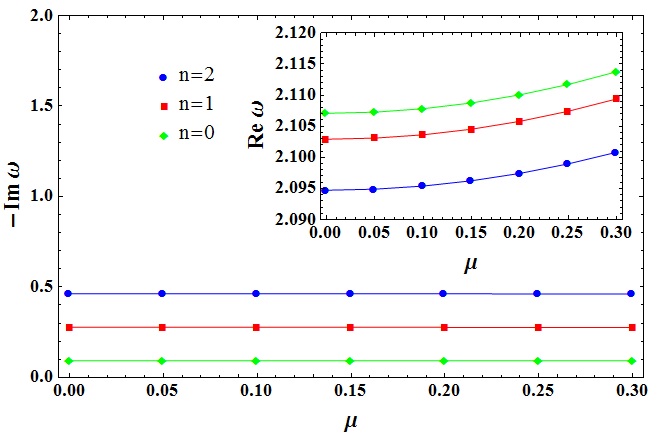}}
\subfigure[]
{\includegraphics[width=.48\textwidth]{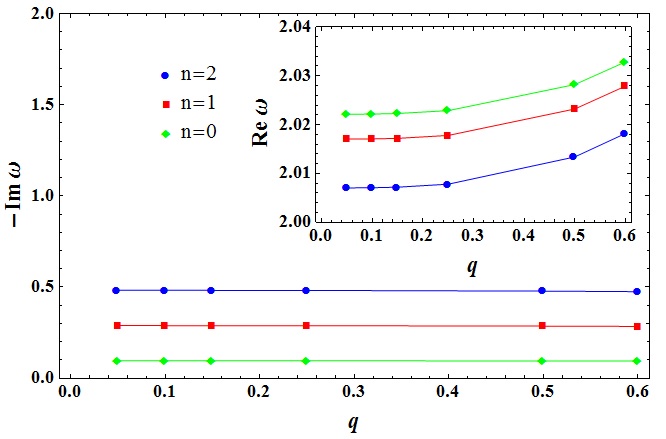}}
	    \caption{{Low-lying QNMs vs. $\mu$ and $q$ for $\xi=20$ and $l=10$.}}
\label{f13}
\end{figure}

Now we investigate how the scalar mass and black hole charge affect the low-lying QNM frequencies. The results are given in Tabs.~(\ref{tab2}) and (\ref{tab3}), respectively and the behaviors of $\omega_{R}$ and $\omega_{I}$ are plotted in Figs.~(\ref{f11})-(\ref{f13}). It is seen from Tab.~(\ref{tab2}) that the higher amounts of $\mu$ correspond to the larger values of $\omega_{R}$ for all three RBHs while for each value of $\mu$ it decreases for higher modes. We have the opposite situation for $\omega_{I}$ as observed in Figs.~(\ref{f11}). These figures are plotted for Hayward black hole and the results are the same for Baedeen and ABG.

 \begin{table}[H] 
 \caption[]{Low-lying QNM frequencies for $\xi=20$, $q=0.5$ and $l=10$. For each value of $n$, the first raw is the 3th order and the second raw is the 6th order of WKB approximation.}
\centering
\begin{tabular}{|c|c|c|c|c|}\hline\hline
 $\mu$ & $n$ & \small{Bardeen} & \small{Hayward} & \small{ABG} \TBstrut \\  \hline
           & $0$   & $2.107094 -i\, 0.092438$  & $2.027410 -i\, 0.095428$   & $2.224729 -i\, 0.091967$  \TBstrut \\
            &   & $2.107101 -i\, 0.092438$  & $2.027417 -i\, 0.095428$   & $2.224736 -i\, 0.091967$ \\ \cline{2-5}
$0.0$   & $1$   & $2.102925 -i\, 0.277561$  & $2.022429 -i\, 0.286585$   & $2.220975 -i\, 0.276108$  \TBstrut \\
&   & $2.102944 -i\, 0.277560$  & $2.022445 -i\, 0.286584$   & $2.220996 -i\, 0.276107$ \\ \cline{2-5}
           & $2$   & $2.094729 -i\, 0.463394$  & $2.012649 -i\, 0.478608$   & $2.213581 -i\, 0.460848$ \TBstrut  \\
           &   & $2.094700 -i\, 0.463413$  & $2.012583 -i\, 0.478636$   & $2.213574 -i\, 0.460860$ \\ \hline
           & $0$   & $2.107280 -i\, 0.092425$  & $2.027614 -i\, 0.095412$   & $2.224896 -i\, 0.091956$  \TBstrut \\
           &   & $2.107287 -i\, 0.092425$  & $2.027621 -i\, 0.095412$   & $2.224903 -i\, 0.091956$ \\ \cline{2-5}
$0.05$ & $1$   & $2.103105 -i\, 0.277521$  & $2.022626 -i\, 0.286539$   & $2.221138 -i\, 0.276074$ \TBstrut \\
&   & $2.103125-i\, 0.277521$  & $2.022642 -i\, 0.286537$   & $2.221159 -i\, 0.276074$ \\ \cline{2-5}
           & $2$   & $2.094899 -i\, 0.463331$  & $2.012832 -i\, 0.478533$   & $2.213735 -i\, 0.460793$ \TBstrut \\
           &   & $2.094869 -i\, 0.463340$  & $2.012767 -i\, 0.478561$   & $2.213728 -i\, 0.460806$ \\ \hline
           & $0$   & $2.107837 -i\, 0.092385$  & $2.028224 -i\, 0.095365$   & $2.225397 -i\, 0.091922$  \TBstrut \\
           &   & $2.107844 -i\, 0.092385$  & $2.028231-i\, 0.095364$   & $2.225403 -i\, 0.091922$ \\ \cline{2-5}
$0.10$   & $1$   & $2.103646 -i\, 0.277403$  & $2.023216 -i\, 0.286399$   & $2.221625 -i\, 0.275974$ \TBstrut \\
&   & $2.103666 -i\, 0.277402$  & $2.023233 -i\, 0.286398$   & $ 2.221646 -i\, 0.275973$ \\ \cline{2-5}
           & $2$   & $2.095409 -i\, 0.463139$  & $2.013384 -i\, 0.478308$   & $2.214198 -i\, 0.460631$ \TBstrut \\
           &   & $2.095378 -i\, 0.463158$  & $2.013318 -i\, 0.478336$   & $2.214191 -i\, 0.460643$ \\ \hline
           & $0$   & $2.110067 -i\, 0.092224$  & $2.030667 -i\, 0.095174$   & $2.227401 -i\, 0.091785$ \TBstrut \\
           &   & $2.110073 -i\, 0.092224$  & $2.030674 -i\, 0.095174$   & $2.227407 -i\, 0.091785$ \\ \cline{2-5}
$0.20$   & $1$   & $2.105811 -i\, 0.276927$  & $2.025578 -i\, 0.285838$   & $2.223576 -i\,  0.275571$ \TBstrut \\
&   & $2.105830 -i\, 0.276927$  & $2.025594 -i\, 0.285836$   & $2.223598 -i\, 0.275570$ \\ \cline{2-5}
           & $2$   & $2.097447 -i\, 0.462374$  & $2.015590 -i\, 0.477408$   & $2.216046 -i\, 0.459981$ \TBstrut \\
           &   & $2.097417 -i\, 0.462393$  & $2.015525 -i\, 0.477436$   & $2.216040 -i\, 0.459993$ \\ \hline
           & $0$   & $2.113785 -i\, 0.091955$  & $2.034740 -i\, 0.094857$   & $2.230742 -i\, 0.091558$ \TBstrut \\
           &   & $2.113792 -i\, 0.091954$  & $2.034747 -i\, 0.094856$   & $2.230749 -i\, 0.091558$ \\ \cline{2-5}
$0.30$   & $1$   & $2.109420 -i\, 0.276134$  & $2.029516 -i\, 0.284902$   & $2.226830 -i\, 0.274899$ \TBstrut \\
&   & $2.109440 -i\, 0.276134$  & $2.029533 -i\, 0.284901$   & $2.226851 -i\, 0.274898$ \\ \cline{2-5}
           & $2$   & $2.100847 -i\, 0.461097$  & $2.019270 -i\, 0.475905$   & $2.219130 -i\, 0.458895$ \TBstrut \\
           &   & $2.100817 -i\, 0.461115$  & $2.019205 -i\, 0.475933$   & $2.219124 -i\, 0.458907$ \\  \hline
\end{tabular}
\label{tab2}
\end{table}
In Tab.~(\ref{tab3}) we list the low-lying QNMs of RBHs for different values of black hole charge $q$. We observe that the higher the value of $q$, the larger the value of real part for particular mode $n$ while the imaginary part decreases when q increases just the same as what happens for the scalar mass. It should be noted that these QNMs are obtained for coupling constant $\xi\!=\!20$ for which the WKB method can be used correctly. In the limit $\xi\rightarrow0$ and by increasing $q$, the imaginary parts increase, which is the same as the behavior of RN black holes presented in \cite{Konoplya:2002ky}.
 \begin{table}[H] 
 \caption[]{Low-lying QNM frequencies for $\xi=20$, $\mu=0.1$ and $l=10$. For each value of $n$, the first raw is the 3th order and the second raw is the 6th order of WKB approximation.}
\centering
\begin{tabular}{|c|c|c|c|c|c|c|}\hline\hline
  $q$ & $n$ & \small{Bardeen} & \small{Hayward} & \small{ABG} \TBstrut \\  \hline
           & $0$   & $2.022876 -i\, 0.096164$  & $2.022141 -i\, 0.096190$   & $2.023720 -i\, 0.096177$  \TBstrut \\
          &    &   $ 2.022882 -i\, 0.096163$  & $2.022146 -i\, 0.096189$   & $2.023726 -i\, 0.096177$ \\ \cline{2-5}
$0.05$ & $1$   & $2.017782 -i\, 0.288811$  & $2.017039 -i\, 0.288889$   & $2.018629 -i\, 0.288851$  \TBstrut \\
        &    &   $2.017792 -i\, 0.288810$  & $2.017048 -i\, 0.288888$   & $ 2.018639 -i\, 0.288849$ \\ \cline{2-5}
           & $2$   & $2.007792 -i\, 0.482376$  & $2.007032 -i\, 0.482509$   & $2.008644 -i\, 0.482442$   \TBstrut \\
          &    &   $2.007701 -i\, 0.482408$  & $2.006940 -i\, 0.482541$   & $2.008552 -i\, 0.482474$ \\ \hline
           & $0$   & $2.025112 -i\, 0.096082$  & $2.022182 -i\, 0.096184$   & $2.028515 -i\, 0.096135$  \TBstrut \\
          &    &   $2.025117 -i\, 0.096082$  & $2.022188 -i\, 0.096184$   & $2.028521 -i\, 0.096134$ \\ \cline{2-5}
$0.10$   & $1$   & $2.020043 -i\, 0.288565$  & $2.017081 -i\, 0.288873$   & $2.023457 -i\, 0.288721$  \TBstrut \\
        &    &   $2.020053 -i\, 0.288563$  & $2.017091 -i\, 0.288872$   & $2.023467 -i\, 0.288720$ \\ \cline{2-5}
           & $2$   & $2.010103 -i\, 0.481960$  & $2.007076 -i\, 0.482481$   & $2.013536 -i\, 0.482219$  \TBstrut \\
           &    &   $2.010012 -i\,0.481992$  & $2.006984 -i\, 0.482513$   & $2.013446 -i\, 0.482251$ \\ \hline
           & $0$   & $2.028877 -i\, 0.095943$  & $2.022295 -i\, 0.096169$   & $2.036640 -i\, 0.096057$  \TBstrut \\
          &    &   $2.028882 -i\, 0.095942$  & $2.022301 -i\, 0.096169$   & $2.036646 -i\, 0.096056$ \\ \cline{2-5}
$0.15$ & $1$   & $2.023851 -i\, 0.288144$  & $2.017196 -i\, 0.288828$   & $2.031638 -i\, 0.288484$  \TBstrut \\
        &    &   $2.023862 -i\, 0.288142$  & $2.017206 -i\, 0.288826$   & $2.031648 -i\, 0.288483$ \\ \cline{2-5}
           & $2$   & $2.013993 -i\, 0.481250$  & $2.007194 -i\, 0.482405$   & $2.021826 -i\, 0.481812$   \TBstrut\\
        &    &   $2.013906 -i\, 0.481281$  & $ 2.007103 -i\, 0.482437$   & $2.021739 -i\, 0.481843$ \\ \hline
           & $0$   & $2.041269 -i\, 0.095466$  & $2.022880 -i\, 0.096091$   & $2.063836 -i\, 0.095743$  \TBstrut \\
           &    &   $2.041275 -i\, 0.095466$  & $2.022885 -i\, 0.096091$   & $2.063842 -i\, 0.095743$ \\ \cline{2-5}
$0.25$ & $1$   & $2.036383 -i\, 0.286703$  & $2.017790 -i\, 0.288593$   & $2.059022 -i\, 0.287530$  \TBstrut \\
        &    &   $2.036394 -i\, 0.286702$  & $2.017800 -i\, 0.288591$   & $2.059034 -i\, 0.287529$ \\ \cline{2-5}
           & $2$   & $2.026796 -i\, 0.478819$  & $2.007806 -i\, 0.482008$   & $2.049575 -i\, 0.480182$  \TBstrut \\
        &    &   $2.026717 -i\, 0.478848$  & $2.007717 -i\, 0.482040$   & $2.049499 -i\, 0.480210$ \\ \hline
           & $0$   & $2.107837 -i\, 0.092385$  & $2.028224 -i\, 0.095365$   & $2.225397 -i\, 0.091922$ \TBstrut \\
           &    &   $2.107844 -i\, 0.092384$  & $2.028231 -i\, 0.095364$   & $2.225403 -i\, 0.091922$ \\ \cline{2-5}
$0.50$   & $1$   & $2.103646 -i\, 0.277403$  & $2.023216 -i\, 0.286399$   & $2.221625 -i\,  0.275974$  \TBstrut \\
        &    &   $ 2.103666 -i\, 0.277402$  & $2.023233 -i\, 0.286397$   & $ 2.221646 -i\, 0.275973$ \\ \cline{2-5}
           & $2$   & $2.095409 -i\, 0.463139$  & $2.013384 -i\, 0.478308$   & $2.214198 -i\, 0.460631$  \TBstrut \\
        &    &   $2.095378 -i\, 0.463158$  & $2.013318 -i\, 0.478336$   & $2.214191 -i\, 0.460643$ \\ \hline
\end{tabular}
\label{tab3}
\end{table}
Fig.~(\ref{f12}) again describes the changes of $\omega_{I}$ and $\omega_{R}$ of fundamental modes as a function of $q$ for Hayward black hole and the behavior of the two other solutions are similar. One can see from Figs.~(\ref{f13}) that for higher modes, by increasing both $\mu$ and $q$ the values of imaginary and real parts of $\omega$, increase and decrease respectively. One can compare the results with the calculations in Refs. \cite{Konoplya:2018qov,Chen:2010qf} which have done for general RN black holes in the presence of coupling the scalar field with Einstein's tensor of the geometry by ignoring the back-reaction effects on the background.

As we emphasized at the beginning of this section, the WKB order might be important in determining accurate values of QNM frequencies such that by increasing the order one obtains better approximation. However, this is not true when we increase the multipole number $l$. It can be inferred from error estimation for each order of WKB formula. This quantity for $\omega_k$, obtained with the WKB formula of the order $k$ for each overtone $n$, is defined as \cite{Konoplya:2019hlu}
\be\label{ee} \Delta_k=\frac{|\omega_{k+1}-\omega_{k-1}|}{2}.\ee
For instance, we have computed this error for Bardeen RBH and the results for $n=1$ are given in Tab.~(\ref{tab4}).

\begin{figure}[H]
\centering
\subfigure[ $l=2$]
{\includegraphics[width=.48\textwidth]{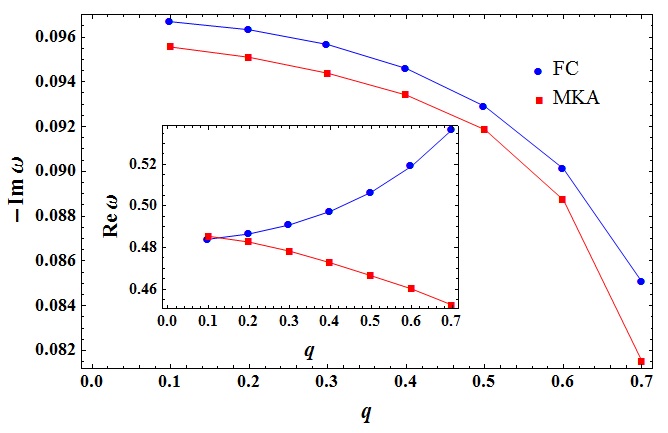}}
\subfigure[$l=10$]
{\includegraphics[width=.485\textwidth]{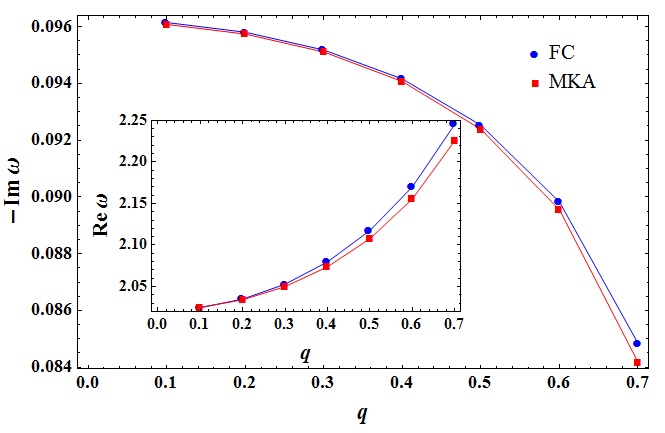}}
	    \caption{{ $-\omega_{I}$ and $\omega_{R}$ of fundamental modes vs. black hole charge $q$.}}
\label{f14}
\end{figure}

\begin{figure}[H]
\centering
\includegraphics[width=9cm,height=6cm]{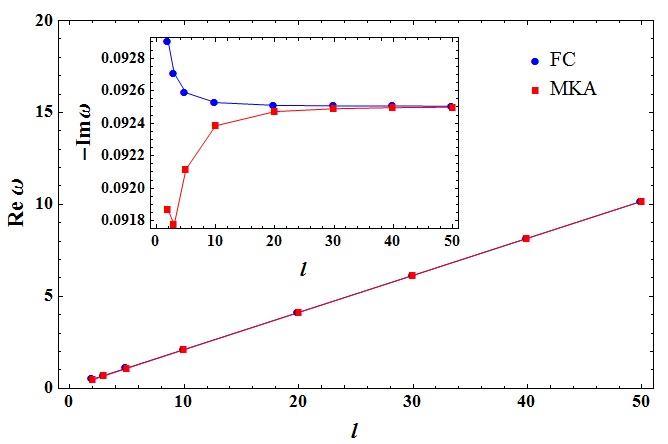}
\caption{$-\omega_{I}$ and $\omega_{R}$ of fundamental modes vs. harmonic number $l$.}
\label{f15}
\end{figure}

For low-lying mode with $l=3$ the error estimation has the lowest value for the 6th order even by increasing the coupling constant which confirms that it is a reliable order of WKB formula. But as seen, by increasing the value of $l$ the approximation becomes more accurate when we consider higher orders. For example, in the case of $l=3$ the best order is 6, for $l=10,50$ is 7, and for $l=100,200$ is 8. It can also be checked that at each order, the increase the value of $l$, the decrease the value of error estimation. Contrary to the monotonical rate of change that occurs in the value of error estimation by increasing $l$, there is no regular change by increasing $\xi$.
 \begin{table}[H] 
 \caption[]{QNMs and error estimation for Bardeen RBH with $q=0.5$ and $\mu=0.1$. }
\centering
\begin{tabular}{|c|c|c|c||c|c|c|}\hline\hline
$l$&$k$&$\omega_k (\xi=60)$&$\Delta_{k}$&$\xi$ &$\omega_k (l=3)$&$\Delta_{k}$\TBstrut\\\hline
  & 3  & $0.604775 -i\, 0.289205$  & $54.8\times 10^{-4}$  &   &  $0.696987 -i\, 0.279126$   & $ 36.8\times 10^{-4}$   \TBstrut\\
  & 4  & $0.604437 -i\, 0.289367$  & $2.1\times 10^{-4}$  &   &  $0.697426 -i\, 0.278950$   &  $2.4\times 10^{-4}$ \\
3& 5  & $0.604345 -i\, 0.289175$  & $1.3\times 10^{-4}$  &0 &  $0.697476 -i\, 0.279075$   &  $0.6\times 10^{-4}$ \\
  & 6  & $0.604190 -i\, 0.289250$  & $1.0\times 10^{-4}$  &   &  $0.697442 -i\, 0.279089$   &  $0.1\times 10^{-4}$ \\
  & 7  & $0.604137 -i\, 0.289139$  & $2.0\times 10^{-4}$  &   &  $0.697437 -i\, 0.279075$   & $ 0.4\times 10^{-4} $\\
  & 8  & $0.603780 -i\, 0.289310$  & $46.5\times 10^{-4}$  &   &  $0.697346 -i\, 0.279112$   & $ 37.1\times 10^{-4}$ \\  \hline
  & 3  & $2.082716 -i\, 0.277022$  & $45.8\times 10^{-5}$  &   &  $0.680563 -i\, 0.278036$   &  $39.9\times 10^{-4}$   \TBstrut\\
  & 4  & $2.082741-i\, 0.277019$  &  $1.2\times 10^{-5}$ &   &  $0.681205 -i\, 0.277774$   &  $3.7\times 10^{-4}$ \\
10& 5  & $2.082741 -i\,0.277023$  & $0.1\times 10^{-5}$  & 10  &  $0.681303 -i\, 0.280431$   & $1.3\times 10^{-4}$  \\
  & 6  & $2.082741 -i\, 0.277023$  & $0.02\times 10^{-5}$  &   &  $0.681213  -i\, 0.278050$   & $0.5\times 10^{-4}$  \\
  & 7  & $2.082741 -i\, 0.277023$  & $0.007\times 10^{-5}$  &   &  $0.681197 -i\, 0.278011$   & $2.3\times 10^{-4}$  \\
  & 8  & $2.082741 -i\, 0.277022$  & $0.05\times 10^{-5}$  &   &  $0.681635 -i\, 0.277832$   &  $174.9\times 10^{-4}$ \\  \hline
   & 3  & $10.174239 -i\, 0.277483 $  & $189.4\times 10^{-7} $ &   &  $0.647784 -i\, 0.278997$   & $ 50.1\times 10^{-4} $  \TBstrut\\
  & 4  & $10.174239 -i\, 0.277483 $  & $ 0.8\times 10^{-7}$ &   &  $0.648649 -i\, 0.278624$   & $ 4.7\times 10^{-4}$ \\
50& 5  & $10.174239 -i\, 0.277483 $  & $0.1\times 10^{-8}$  & 30 &  $0.648662 -i\, 0.278655$   & $1.1\times 10^{-4}$  \\
  & 6  & $10.174239 -i\, 0.277483 $  & $0.3\times 10^{-10}$  &   &  $0.648864  -i\, 0.278568$   & $1.3\times 10^{-4}$  \\
  & 7  & $10.174239 -i\, 0.277483 $  &$0.1\times 10^{-11} $ &   &  $0.648922 -i\, 0.278703$   & $2.3\times 10^{-4}$  \\
  & 8  & $10.174239 -i\, 0.277483 $  &  $0.2\times 10^{-11}$ &   &  $0.648518 -i\, 0.278876$   & $160.8\times 10^{-4}$ \\  \hline
  & 3  & $20.258360 -i\, 0.277506$  & $4.77\times 10^{-6}$  &   &  $0.604776 -i\, 0.289206$   & $54.8\times 10^{-4}$   \TBstrut\\
  & 4  & $20.258360 -i\, 0.277506$  &$0.1\times 10^{-7} $ &   &  $0.604438 -i\, 0.289368$   & $ 2.1\times 10^{-4}$ \\
100& 5  & $20.258360 -i\, 0.277506$  & $ 0.1\times 10^{-9}$ &  60 &  $0.604346 -i\, 0.289176$   & $ 1.3\times 10^{-4}$ \\
  & 6  & $20.258360 -i\, 0.277506$  & $0.1\times 10^{-11}$  &   &  $0.604190  -i\, 0.289250$   & $ 1.0\times 10^{-4}$ \\
  & 7  & $20.258360 -i\, 0.277506$  & $0.2 \times 10^{-13}$ &   &  $0.604137 -i\, 0.289140$   & $ 2.1\times 10^{-4}$ \\
  & 8  & $20.258360 -i\, 0.277506$  & $0.4 \times 10^{-14}$ &   &  $0.603781 -i\, 0.289310$   &$  46.5\times 10^{-4}$ \\  \hline
  & 3  & $40.421291 -i\, 0.277512$  &  $1.2\times 10^{-6}$ &   &  $0.571613 -i\, 0.302528$   & $ 51.5\times 10^{-4}$   \TBstrut\\
  & 4  & $40.421291 -i\, 0.277512$  & $0.1\times 10^{-8}$  &   &  $0.570888 -i\, 0.302912$   &$ 4.2\times 10^{-4}$  \\
200& 5  & $40.421291 -i\, 0.277512$  & $0.6\times 10^{-11}$  & 90  &  $0.570994 -i\, 0.303112$   & $1.1\times 10^{-4} $ \\
  & 6  & $40.421291 -i\, 0.277512$  & $0.3\times 10^{-13}$  &   &  $0.570942  -i\, 0.303140$   &  $0.3\times 10^{-4}$ \\
  & 7  & $40.421291 -i\, 0.277512$  &  $0.1\times 10^{-15}$ &   &  $0.570965  -i\, 0.303183$   &$ 0.2\times 10^{-4}$  \\
  & 8  & $40.421291 -i\, 0.277512$  & $0.0$  &   &  $0.570963 -i\, 0.303184$   & $ 1.0\times 10^{-4}$ \\ \hline
\end{tabular}
\label{tab4}
\end{table}
Finally, we compare our results with the results obtained by Fernando and Correa \cite{Fernando:2012yw} who studied QNMs of the Bardeen black hole due to the scalar perturbations with $\mu\!=\!0$ and in zero coupling limit $\xi\!=\!0$. Our results for Bardeen black hole when $\xi\!=\!20, \, \mu\!=\!0.1$  in the case of fundamental modes ($n\!=\!0$) are given in Figs.~(\ref{f14},\ref{f15}). As one can deduce from the plots in Figs.~(\ref{f14}a), the behavior of imaginary parts of the two models for $l\!=\!2$ are similar but not exactly the same while the real parts have totally different behavior. However, Figs.~(\ref{f14}b) shows that for $l\!=\!10$, that is, when we increase $l$, the modes nearly coincide with each other.

The comparison of different values of multipole number $l$ in Figs.~(\ref{f15}) display explicitly that the real parts are exactly the same which indicates that they might be independent of $\xi$ and $\mu$ while the imaginary parts have different behavior in lower $l$ but for large values as noticed previously, they become equal and tend to an asymptotic value in compatible with eikonal limit.
\subsection{GFs of RBHs in WKB approximation}
\begin{figure}[H]
\centering
\subfigure[Bardeen]
{\includegraphics[width=.42\textwidth,height=4.5cm]{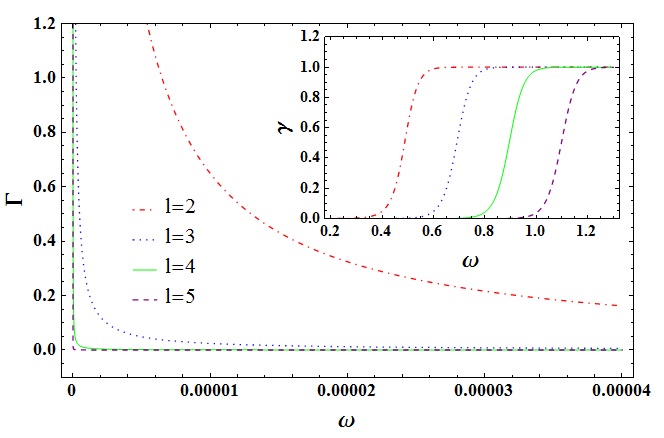}}
\subfigure[Hayward]
{\includegraphics[width=.42\textwidth,height=4.5cm]{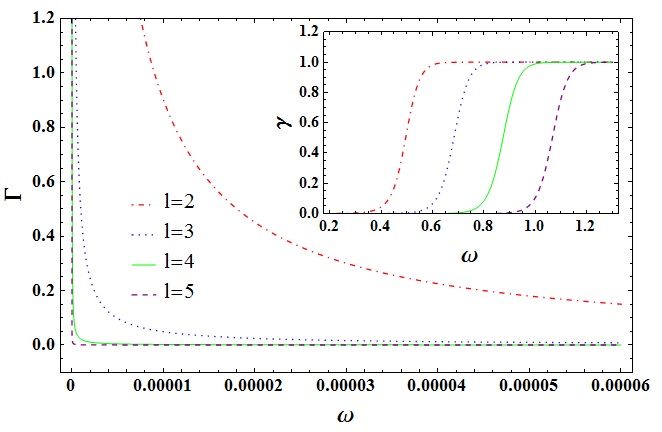}}
\subfigure[ABG]
{\includegraphics[width=.42\textwidth,height=4.5cm]{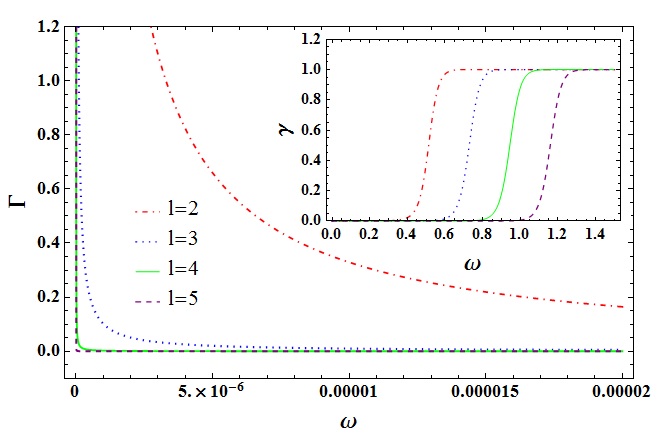}}
	    \caption{\small{HRs and GFs for different $l$ when $\mu=0.1$, $ q=0.5$ and $\xi=20$.}}
\label{f16}
\end{figure}

\begin{figure}[H]
\centering
\subfigure[Bardeen]
{\includegraphics[width=.42\textwidth,height=4.5cm]{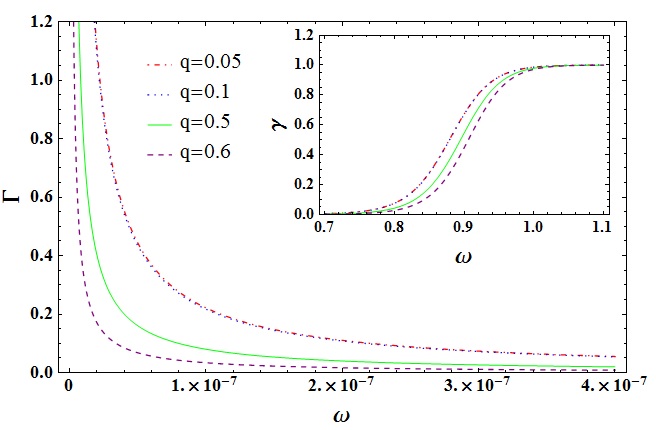}}
\subfigure[Hayward]
{\includegraphics[width=.42\textwidth,height=4.5cm]{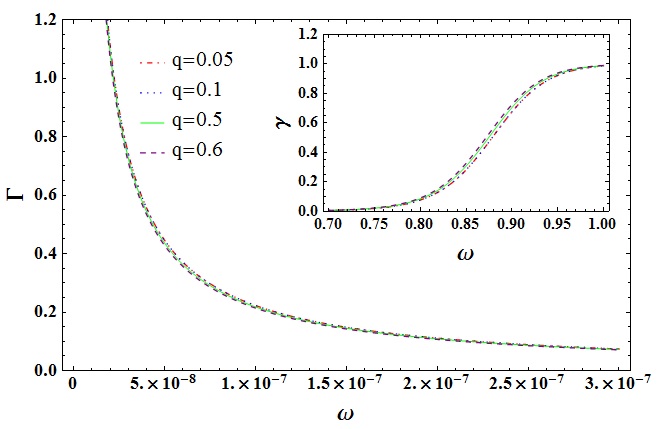}}
\subfigure[ABG]
{\includegraphics[width=.42\textwidth,height=4.5cm]{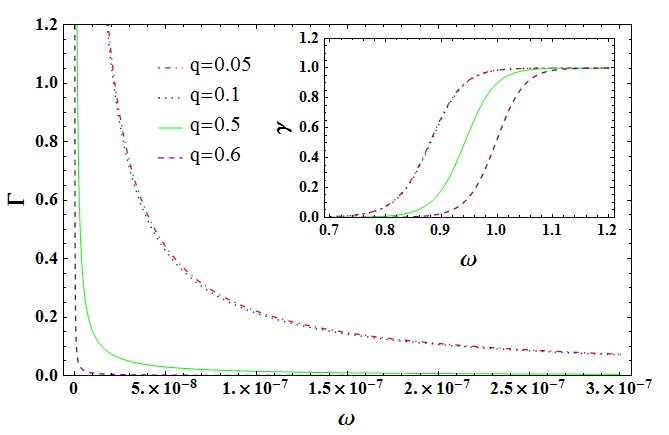}}
	    \caption{\small{HRs and GFs for different $q$ when $\mu=0.1$, $l=4$ and $\xi=20$.}}
\label{f17}
\end{figure}

\begin{figure}[H]
\centering
\subfigure[Bardeen]
{\includegraphics[width=.42\textwidth,height=4.2cm]{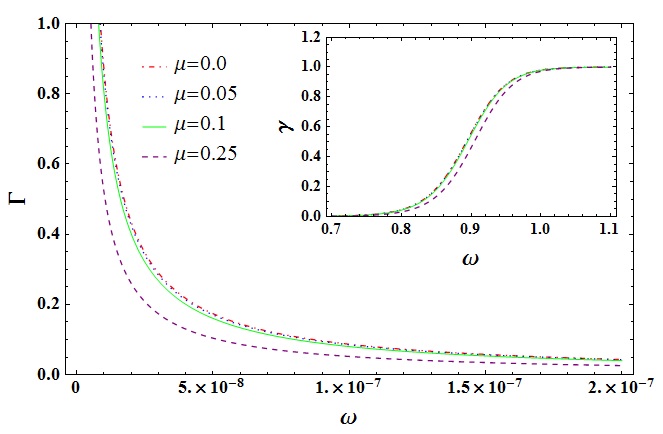}}
\subfigure[Hayward]
{\includegraphics[width=.42\textwidth,height=4.2cm]{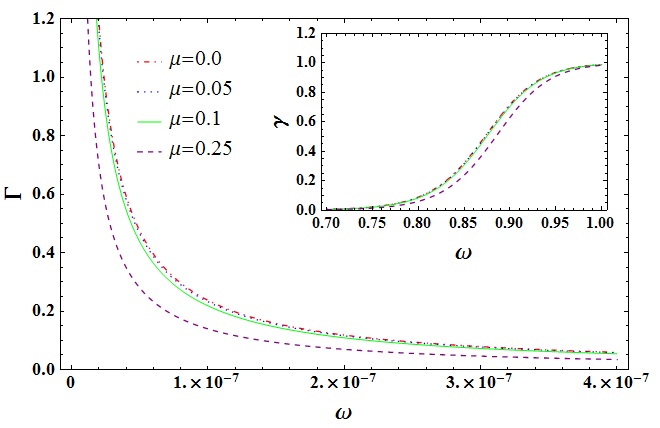}}
\subfigure[ABG]
{\includegraphics[width=.42\textwidth,height=4.2cm]{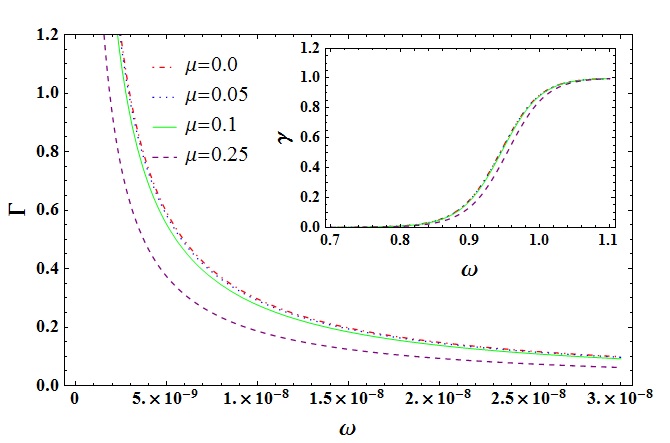}}
	    \caption{\small{HRs and GFs for different $\mu$ when $\xi=20$, $l=4$ and $ q=0.5$.}}
\label{f18}
\end{figure}

\begin{figure}[H]
\centering
\subfigure[Bardeen]
{\includegraphics[width=.42\textwidth,height=4.2cm]{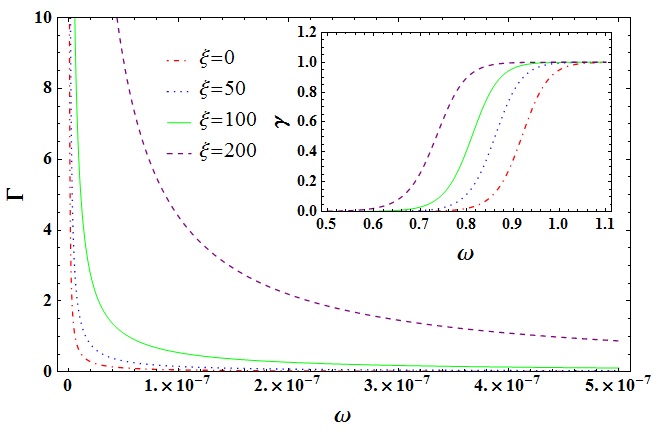}}
\subfigure[Hayward]
{\includegraphics[width=.42\textwidth,height=4.2cm]{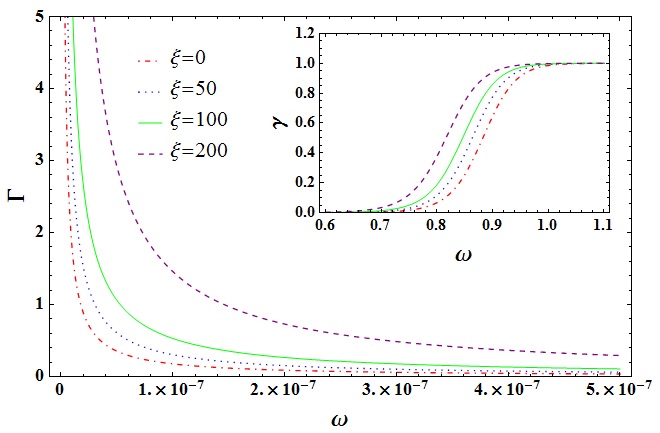}}
\subfigure[ABG]
{\includegraphics[width=.42\textwidth,height=4.2cm]{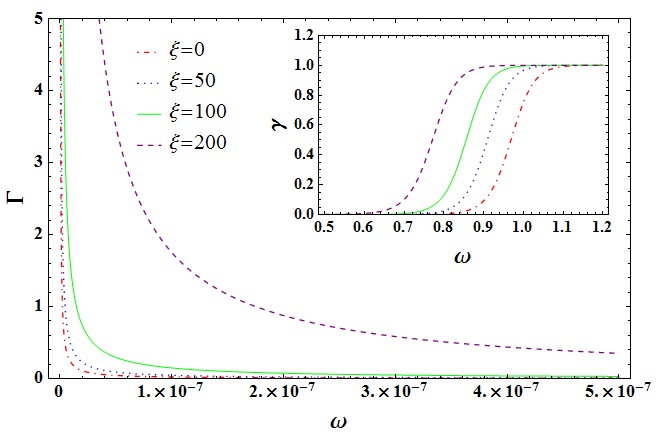}}
	    \caption{\small{HRs and GFs for different $\xi$ when $\mu=0.1$, $l=4$ and $ q=0.5$.}}
\label{f19}
\end{figure}

The emission rate of a black hole in a mode with frequency $\omega$ is given by 
\be\label{HR1} \Gamma(\omega)=\frac{1}{e^{\beta\omega}\pm1}\frac{d^3k}{(2\pi)^3}\,,\ee
where $\beta$ is the inverse of the Hawking temperature of the black hole at event horizon and the minus (plus) sign is used when considering bosons (fermions) \cite{Hawking:1974sw,Hawking:1974rv}. The geometry outside the event horizon is non-trivial and acts as a potential barrier for the Hawking radiation (HR) from the black hole. Part of it will be transmitted and will travel freely to infinity, whereas the rest will be reflected back into the black hole. The radiation recorded by the distant observer will no longer appear as a black body. So, the spectrum of HR of a black hole measured by an observer at infinity for a frequency mode $\omega$ is
\be\label{HR2} \Gamma(\omega)=\frac{\gamma_{l}(\omega)}{e^{\beta\omega}\pm1}\frac{d^3k}{(2\pi)^3},\ee
where $\gamma$ is the GF of the angular quantum number $l$. In general for the wave-function of a particle propagating around a black hole we have \cite{Iyer:1986np}
\bea \label{wf} \psi (r_{*}) &\!\!\!\!=\!\!\!\!&T(\omega)e^{-i\omega r_{*}}, \qquad r_{*}\rightarrow -\infty \nn\\
 \psi(r_{*}) &\!\!\!\!=\!\!\!\!& e^{-i\omega r_{*}}+R(\omega)e^{+i\omega r_{*}}, \qquad r_{*}\rightarrow +\infty \eea
where $R(\omega)$ and $T(\omega)$ are the reflection and transmission coefficients, respectively, and they are related to each other by $|R|^2+|T|^2\!=\!1$. So, in general the GF is defined as
\be\label{GF1} \gamma_{l}(\omega)=|T(\omega)|^2. \ee
The reflection and transmission coefficients in WKB approximation are given in Ref.~\cite{Konoplya:2009hv}. Since in WKB approximation the value with highest accuracy is obtained for $\omega^2\sim V_{\circ}$, we calculate the GFs in this limit.
The numerical results of the HRs and GFs as a function of $\omega$ for different values of parameters are summarized in Figs.~(\ref{f16})-(\ref{f19}). 
As is obvious from the figures, the GF $\gamma(\omega)$ or equivalently the transmission coefficient $|T(\omega)|^2$ goes to zero in the limit $\omega\rightarrow 0$ and to one for large $\omega$ as expected. From Figs.~(\ref{f16}) we see that HR decreases as $l$ increases for all RBHs and transmission occurs for larger $\omega$. In addition Figs.~(\ref{f17}) shows that these factors behave similarly, i.e. HR decreases with $\omega$ as the black hole charge $q$ increases. Our results are consistent with the calculations that have been done in \cite{Chakrabarty:2018skk,Dey:2018cws} for Bardeen de-Sitter black holes with other physical fields but in the absence of any mass and coupling term.

The other two important parameters that have not been considered before are the scalar mass $\mu$ and coupling $\xi$ for RBHs spacetimes. From Figs.~(\ref{f18}) one can see that by increasing $\mu$, the HRs and GFs have the same behavior with respect to the parameters $l$ and $q$ while in the case of coupling an opposite situation occurs in Figs.~(\ref{f19}), that is, when we increase $\xi$ the HR of RBH increases with $\omega$ but the GF decreases.

As discussed in Sec.~3, for $\xi>200$ the effective potential has a well outside the event horizon, so by plotting HRs and GFs in this regime we observe that GFs are negative which is an illogical result. This phenomenon, when the reflection coefficient $|R|$ can be larger than unity, is called superradiance \cite{Brito:2015oca,Konoplya:2019hlu}. As is obvious, the reflection coefficient cannot be larger than unity for any imaginary $\kappa$, so superradiance cannot be described by WKB formula at least when the effective potential is real. We encountered similar description for calculating the QNMs from WKB method in Sec.~3.
\subsection{AQ of RBHs from near horizon approximation}
The AQ of a black hole horizon in units of Planck length is a long-standing issue, that goes back to the profound revelations of Bekenstein in the early seventies \cite{Bekenstein:1973ur}. Most significantly, the area of the horizon was shown to be an adiabatic invariant \cite{Bekenstein:1974jk}. According to the Ehrenfest’s principle
it has a discrete and evenly spaced spectrum
\be A_{n}=\epsilon \hbar\cdot n=\epsilon l_{p}^2\cdot n,\ee
where $ l_{p}$ is the Planck length, $A_{n}$ denotes the area spectrum and $n$ is the quantum number.
Bekenstein professed that $\epsilon\!=\! 8\pi$ and the horizon is formed by patches of equal area $\epsilon \hbar$. An important step in this direction was made by Hod \cite{Hod:1998vk}. He suggested that the spacing $\epsilon \hbar$ of the area spectrum can be determined by utilizing the QNM frequencies of an oscillating black hole.

On the other hand, in \cite{Kunstatter:2002pj} Kunstatter pointed out that, for a system with energy $E$ and vibrational frequency $\omega(E)$, a natural adiabatic invariant quantity is
\be\label{ai} I=\int\frac{dE}{\Delta\omega(E)}=\int\frac{T_H dS}{\Delta\omega},\ee
where $T_H$ and $S$ are the temperature and entropy of the black hole, respectively.  In the large $n$ limit, this adiabatic invariant is quantized according to the Bohr’s correspondence principle in the form
\be\label{BC} I\sim n\hbar.\ee
If one interpret the vibrational frequency $\Delta\omega(E)$ as the real part of QNM frequencies and replacing the energy $E$ with the black hole mass $M$,
 then using the Bohr-Sommerfeld quantization condition (\ref{BC}) at large $n$ limit, the AQ of the Schwarzschild black hole is calculated as
\be\label{area} A_{n}=4\hbar \ln3\cdot n, \ee
which is consistent with Hod’s result (see e.g \cite{Dreyer:2002vy}).

To avoid several problems in the interpretation of QNM frequencies when compared with macroscopical systems, Maggiore proposed \cite{Maggiore:2007nq} that one can treat a perturbed black hole as a damped harmonic oscillator with the quasinormal normal frequency $\omega_0\!=\!\sqrt{\omega_{R} ^2+\omega_{I} ^2}$ such that for long-lived QNM, i.e. $\omega_{I}\rightarrow 0$, the frequency of the harmonic oscillator becomes $\omega_\circ\!=\!\omega_{R}$. However, the most interesting case is that of highly excited QNMs for which $\omega_{I}\gg \omega_{R}$ then the frequency of the harmonic oscillator becomes $\omega_\circ\!=\!\omega_{I}$. So, one has to employ $\omega_I$ rather than $\omega_{R}$, because in order to derive the quantum spectrum of a black hole using its QNMs, the black hole has to be treated as a collection of damped harmonic oscillators. Since we are interested in highly excited black holes, i.e. large $n$, the proper frequency is now $\omega_\circ\!=\!\omega_{I}$. Therefore, the small variations in the mass of the black hole stem from the transition frequency $\Delta \omega(E)\!=\!|\omega_{I}|_{n}-|\omega_{I}|_{n-1}$ \cite{Vagenas:2008yi}.

Even though there is not an exact solution to the radial wave equation (\ref{req}) but in order to find the QNMs analytically, we consider this equation in the near horizon region of RBHs. As denoted in Figs~(\ref{f4}), by increasing the quantum harmonic number $l$, the hight of the potential barrier will increases such that for large $l$ we have an effective potential and the scalar waves are confined between the horizon and the potential peak. In other words, this will yield to characteristic resonance modes of the confined scalar fields in the geometry of RBHs. To this end, the scalar field is imposed to be terminated at the peak and to be purely ingoing wave at the horizon and the QNMs are computed using the poles of the scattering amplitude in the Born approximation, namely
\be\psi \sim\Big\{\begin{array}{cc}
e^{i\omega{r^{*}}} & r^{*} \rightarrow -\infty \nn\\
0& potential\, peak.
\end{array}\ee

Near the event horizon the metric function $f (r)$ in (\ref{bm}) can be expanded as follows
\be\label{nhf}f(r)\simeq f'(r_h)(r-r_h)+\frac{ f''(r_h)}{2}(r-r_h)^2+\dots,\ee
where $r_h$ is the event horizon of RBH. By defining the new variable $x\!=\!r-r_h$ and using the fact that the surface gravity at the horizon is given by $\kappa\!=\!f'(r_h)/2$, we have
\be\label{newh} f(x)=2\kappa x+\frac{f''(r_h)}{2} x^2+\mathcal{O}(x^3).\ee
Substituting this function in potential (\ref{veff}) it can be written as
\be\label{nhv} V(x)\simeq a x^2+b x,\ee
where
\bea \label{potcons} a&\!\!\!\!\!=\!\!\!\!\!&-\frac{8\kappa (\kappa r_h+2(1-\kappa r_h )\xi+l(l+1))}{2r_h^3}\nn\\
&\!+\!&\frac{r_h f''(r_h)\left(6\kappa r_h+(2-24\kappa r_h)\xi+l(l+1)+r_h^2 \mu^2-r_h^2 \xi f''(r_h)\right)}{2r_h^3},\nn\\
b&\!\!\!\!\!=\!\!\!\!\!&\frac{2\kappa\left(2\kappa r_h +(2-8\kappa r_h )\xi+l(l+1)+r_h^2 \mu^2\right)}{r_h^2}\!-\!2\kappa\,\xi f''(r_h).
\eea
Thus, according to tortoise coordinate (\ref{tc}) in this limit, i.e. $r^{*}\!\simeq \!\frac{1}{2\kappa} \ln{x}$, the near horizon of the one-dimensional Schr\"{o}dinger equation (\ref{sceq}) recasts as
\be\label{scwit} 4\kappa^2 x^2\frac{d^2 \psi}{dx^2}+4\kappa^2 x\frac{d\psi}{dx}+(\omega^2-V(x))\psi(x)=0.\ee
This equation can be solved analytically in terms of Hypergeometric and Laguerr functions. Then, by transforming the Laguerr function in terms of Hypergeometric ${_1} F_1$ we have
\be\label{hgf} \psi(x)\sim D_1 x^{\tfrac{i\omega}{2\kappa}} U\left[A,B,Cx\right]+D_2\,\frac{\Gamma\left[A\right]}{\Gamma\left[1+A-B\right]\Gamma\left[B\right]}\, x^{\tfrac{i\omega}{2\kappa}} {_1} F_1\left[A,B,Cx\right],\ee
where $D_1$ and $D_2$ are the constants of integration and three new constants are defined as
\be\label{conshg} A=\frac{b}{4\sqrt{a}\kappa}+\frac12+i\frac{\omega}{2\kappa},\quad B=1+i\frac{\omega}{\kappa},\quad C=\frac{\sqrt{a}}{\kappa}.\ee

Now, taking the near horizon limit $x\ll1$ we obtain
\be\label{nhsol} \psi\sim D_1 \frac{\Gamma\left[B-1\right]}{\Gamma \left[A\right]}\,x^{-\tfrac{i\omega}{2\kappa}}+\left(D_1 \frac{\Gamma\left[1-B\right]}{\Gamma\left[1+A-B\right]}+D_2 \frac{\Gamma\left[A\right]}{\Gamma\left[1+A-B\right] \Gamma\left[B\right]}\right)\,x^{\tfrac{i\omega}{2\kappa}}.\ee
The boundary condition that the wave function of particle be purely ingoing at the horizon demands the first term, which represents an outgoing wave, must be vanished. From the properties of the Gamma functions in general textbooks
of Mathematics (Weierstras’s form) \cite{AS} and due to vanishing the first term, we should have $A\!=\!-n$ where $n$ is a non-negative integer number called overtone number. Hence we are able to determine the QNM frequencies for RBHs as
\be\label{qnmnh1} \omega_n \simeq \frac{b}{2\sqrt{a}}+i\kappa (2n+1), \ee
where the constants $a$ and $b$ are given by (\ref{potcons}).

For the highly excited states we have $\omega_{I}\gg\omega_{R}$, therefore in this limit we have
\be \Delta\omega\approx |\omega_{I}|_{n}-|\omega_{I}|_{n-1}=2\kappa=4\pi T_H.\ee
Substituting this into Eq.~(\ref{ai}) the adiabatic invariant becomes
\be\label{ai2} I=\frac{S}{4\pi},\ee
where due to the Bohr-Sommerfeld quantization condition (\ref{BC}), the entropy is discrete and equidistant as
\be S_n=4\pi n.\ee
Since the Bekenstein-Hawking entropy of black holes is equal to $\frac{A}{4}$ with $\hbar\!=\!G\!=\!1$, then the AQ of the three RBHs are the same and is given by
\be \label{aq} A=16\pi n.\ee
The result shows that the area spectrum in Einstein gravity, even in the presence of matter fields, is equally spaced and is consistent with the proposal in \cite{Kothawala:2008in} which claims that the black holes in Einstein’s theories should have equidistant area spectrum.
\subsection{Shadow radius of RBHs}
There is a deep connection between the QNM and SR of a black hole that has attracted much attention in recent years as the first image of the black hole was being released by EHT \cite{Akiyama:2019cqa,Akiyama:2019eap}. In this subsection we consider the correspondence between the SR and the real part of the QNM. Black holes cast shadow on the bright background as an optical appearance due to the strong gravitational lensing effect. On the other hand, Geodesic motion determines important features of spacetime. Null unstable geodesics are closely related to the appearance of compact objects to external observers and have been associated with the characteristic modes of black holes. In this regard, it has been shown in ref.~\cite{Cardoso:2008bp} that in the eikonal limit, the real part of the QNM is related to the angular velocity of the unstable null geodesic. However, the imaginary part is associated to the Lyapunov exponent that determines the instability time scale of the orbits. That is,
\be\label{sqnm1} \omega_n=\Omega l-i \left(n+\frac12\right)|\lambda|,\ee
where $\Omega$ is the angular velocity at the unstable null geodesic and $\lambda$ denotes the Lyapunov exponent \cite{Cardoso:2008bp}. This result is independent of the field equations and only assumes a stationary, spherically symmetric and asymptotically flat line element.

The shadow around a black hole as seen by a distant observer is described by the radius $R_s$
\be\label{sr1} \omega_R=\lim_{l\gg1}\frac{l}{R_s},\ee
Hence, we can quickly rewrite QNM frequencies in (\ref{sqnm1}) as follows
\be\label{sqnm2} \omega_n=\lim_{l\gg1}\frac{l}{R_s}-i \left(n+\frac12\right)|\lambda|,\ee
On the other side, for spherically symmetric geometries, the SR of the black hole is related to the photon sphere as given by \cite{Shaikh:2018lcc}
\be\label{sr2} R_s=\frac{r}{\sqrt{f(r)}}\Big|_{r=r_{ps}},\ee
where the photon sphere $r_{ps}$ is the solution of equation $2f(r)\!-\!rf'(r)\!=\!0$ at the eikonal limit \cite{Cardoso:2008bp}. The unstable photon orbits constitute the photon sphere, and they define the boundary of the shadow cast by a compact object. Photons from a distant source with impact parameter larger than
the critical impact parameter remain outside the photon sphere and reach the observer. However, photons with impact parameters smaller than the critical impact parameter are captured within the photon sphere and do not reach the observer, thereby creating dark spots in the observer’s sky. The union of these dark spots constitutes the shadow. Therefore, the apparent shape of the shadow projected in the observer’s sky is a circular disk whose radius is given by the critical impact parameter $r_{ps}=3\sqrt{3}M$ for the Schwarzschild black hole.

By expanding the real part of the QNM frequencies (\ref{qnm}) in the eikonal limit, i.e. large $l$, we can find the SR of the black holes. So, up to the sub-leading order for Bardeen, Hayward and ABG black holes we have respectively,
\be\label{bsr} \omega_R=\frac{1}{3\sqrt{3}M}(l+\frac12)\left(1+\frac{q^2}{6M^2}\right)+\mathcal{O}(l^{-1}),\ee
\be\label{hsr} \omega_R=\frac{1}{3\sqrt{3}M}(l+\frac12)\left(1+\frac{q^3}{27M^3}\right)+\mathcal{O}(l^{-1}),\ee
\be\label{abgsr} \omega_R=\frac{1}{3\sqrt{3}M}(l+\frac12)\left(1+\frac{q^2}{3M^2}\right)+\mathcal{O}(l^{-1}).\ee
As is obvious these results are consistent with the SR of Schwarzschild black hole in the limit $q\rightarrow 0$. It is then clear that the SR should decrease by increasing the magnitude of the electric charge $q$. For the sake of comparison with the SR of null geodesic instability we calculate the SR using (\ref{sr1}) and (\ref{sr2}) which their numerical values are given in Tab.~(\ref{tab5}). As seen in each column the value of $R_s$ decreases when we increase the value of $q$. Also, the numerical values of SR are the same with a good accuracy for two approaches.
 \begin{table}[H] 
 \caption[]{SR from null geodesic (\ref{sr2}) and real part of QNMs (\ref{sr1}) for $M=1$.}
\centering
\begin{tabular}{|c|c|c|c|c|c|c|}\hline\hline
  $q$ & $R_{s}$({\footnotesize{Bardeen}})  & $\frac{l}{\omega_{R}}$(\footnotesize{Bardeen}) & $R_{s}$(\footnotesize{Hayward}) & $\frac{l}{\omega_{R}}$(\footnotesize{Hayward}) & $R_{s}$(\footnotesize{ABG}) & $\frac{l}{\omega_{R}}$(\footnotesize{ABG}) \TBstrut \\  \hline
    $0.01$   & $5.196066$   & $5.196065$  & $5.196152$   & $5.196149$ & $5.195979$ & $5.195979$  \TBstrut \\
$0.05$ & $5.193986$   & $5.193988$  & $5.196128$   & $5.196125$ & $5.191820$ & $5.191825$  \\
$0.1$   & $5.187465$   & $5.187506$  & $5.195959$   & $5.195957$ & $5.178754$ & $5.178890$  \\
 $0.2$  & $5.161077$   & $5.161741$  & $5.194611$   & $5.194610$ & $5.125615$ & $5.127782$  \\ [0.5ex]\hline
\end{tabular}
\label{tab5}
\end{table}
\section{Conclusions}
In this paper we performed a numerical study for QNMs of a neutral massive scalar field perturbations which is non-minimally coupled to the curvature of the spacetime geometry around some singularity free black holes known as RBHs. That is the case of great interest in the cosmological and modified gravity models. We have studied the most general examples of such spherically symmetric RBHs such as Bardeen, Hayward, and ABG black holes. We have also considered the GFs, AQ and SR as three different features of RBHs which are closely related to the QNM frequencies of their perturbations.

The scalar field equation in these geometries recast in a Schr\"{o}dinger-like equation with an effective potential. We have investigated the dependence of this potential on the physical parameters of RBHs in Figs.~(\ref{f2}-\ref{f5}). By plotting this potential as a function of $r$, we observed that it has a potential barrier with a finite maximum value which depends on different values of $\mu,\,l,\,\xi,\,$ and $q$. By analysing this potential we noticed that due to the appearance of a negative potential in a certain region we can predict the possibility of finding instabilities for some ranges of parameters, for instance in the large coupling constants illustrated in Fig.~(\ref{f3}). As the main purpose of this paper, we have calculated the extended spectrum of the QNMs of a test scalar field perturbations using the 3th and 6th order of WKB approximation. The results of the calculations are given in Tabs.~(\ref{tab1}-\ref{tab4}) and Figs.~(\ref{f6}-\ref{f15}). Although we did not calculate the fundamental modes ($l\! =\! n\! =\! 0$), it does not mean that they cannot be existed. One can get this modes by other approaches. So, we only considered the low-lying modes with $l>n$. It was shown that the frequencies all have a negative imaginary part, which means that the propagation of scalar field is stable in this background.

A summary of the results is as follows;
\begin{itemize}
\item The rate of normal oscillations, $\omega_{R}$, increases when the multipole number $l$ increases, and the same fact for the scalar field mass $\mu$ and black hole charge $q$, while it decreases when coupling $\xi$ increases. In other words, for large couplings the scalar field is bound to spacetime geometry.
\item Since the scalar field is a test particle, the values of its mass $\mu$ cannot be large in comparison to the black hole mass. Therefore the effects of the mass of the scalar field does not render a significant change in the QNMs spectrum. Also, for small values of the scalar field coupling as expected, the coupling does not affect much the response to the perturbing field. In fact, it is needed large values of the coupling constant to make a perceptible difference in the results. However, as an important point, increasing the value of $l$ reduces the effectiveness of $\xi$ and $\mu$ on the QNM frequencies.
\item Though the increase of the mass causes the rate of damping or dissipating energy of gravitational waves decreases, but it was shown that we have different behavior for each black hole by increasing the coupling. However, in most of the cases $\omega_{I}$ decreases which means that the timescale for approaching to a thermal equilibrium increases for the modes by changing the parameter in the context of AdS/CFT.
\item The consideration of error estimation for different orders of WKB formula in Tab.~(\ref{tab4}) showed that in the case of low-lying modes, the 6th-order is a reliable order of WKB formula only for small $l$ and $\xi$ such that by increasing the value of $l$, the approximation becomes more accurate when we consider higher orders.
\item The rate of emission is suppressed with respect to $\omega$ as we increase $\mu$, $l$, and $q$ while amplified by increasing $\xi$. This fact states that the strength of the coupling constant in scalar-tensor theories play an important role in studying the stability and thermodynamics of the black holes.
\item The AQ of RBHs is calculated from an adiabatic invariant quantization and in the case of highly excited modes where the spectrum is given in (\ref{aq}). The SR of RBH is determined from the real part of QNMs in the eikonal limit which the corresponding expressions are determined from (\ref{bsr})-(\ref{abgsr}). The results show that by increasing the value of charge $q$ the SR will be smaller. The numerical values of SR in Tab.~(\ref{tab5}) show that the results are compatible with the ones obtained from null geodesic condition (\ref{sr2}).
\end{itemize}

As for future works, it remains to study whether the WKB method can be employed to study QNMs of other kinds of non-minimal couplings in scalar-tensor theories or not. Also, there are still some issues for other regular black holes as in Refs. \cite{Nicolini:2005vd,Ansoldi:2006vg,Nicolini:2009gw}, but we hope to report on more general results for this in the future.

\section*{Acknowledgment}
The authors, specially M.K. would like to thank R. Konoplya and V. Cardoso for useful discussion. We would also like to acknowledge J. Matyjasek and M. Opala \cite{Matyjasek:2017psv} for sharing their Mathematica$\textregistered$  notebook with higher order WKB corrections.


\end{document}